\documentclass[fleqn,usenatbib]{mnras}

\usepackage[T1]{fontenc}
\usepackage{ae,aecompl}

\usepackage{graphicx}	
\usepackage{amsmath}	
\usepackage{amssymb}	
\usepackage{newtxtext,newtxmath}

\title[ASAS-SN Catalog of Variable Stars V]{The ASAS-SN Catalog of Variable Stars V: \textit{Variables in the Southern Hemisphere}}

\author[T. Jayasinghe et al.]{T. Jayasinghe$^{1,2}$\thanks{E-mail: jayasinghearachchilage.1@osu.edu},
K. Z. Stanek$^{1,2}$,
C. S. Kochanek$^{1,2}$,
B. J. Shappee$^{3}$,
\newauthor 
T. W. -S. Holoien$^{4}$,
Todd A. Thompson$^{1,2,5}$,
J. L. Prieto$^{6,7}$,
Subo Dong$^{8}$,
M. Pawlak$^{9}$,
\newauthor 
O. Pejcha$^{9}$,
J. V. Shields$^{1}$,
G. Pojmanski$^{10}$,
S. Otero$^{11}$,
N. Hurst$^{12}$,
C. A. Britt$^{12}$,
\newauthor 
D. Will$^{1,12}$
\\
$^{1}$Department of Astronomy, The Ohio State University, 140 West 18th Avenue, Columbus, OH 43210, USA\\
$^{2}$Center for Cosmology and Astroparticle Physics, The Ohio State University, 191 W. Woodruff Avenue, Columbus, OH 43210, USA\\
$^{3}$Institute for Astronomy, University of Hawaii, 2680 Woodlawn Drive, Honolulu, HI 96822,USA\\
$^{4}$Carnegie Observatories, 813 Santa Barbara Street, Pasadena, CA 91101, USA\\
$^{5}$Institute for Advanced Study, Princeton, NJ, 08540\\
$^{6}$N\'ucleo de Astronom\'ia de la Facultad de Ingenier\'ia y Ciencias, Universidad Diego Portales, Av. Ej\'ercito 441, Santiago, Chile\\
$^{7}$Millennium Institute of Astrophysics, Santiago, Chile\\
$^{8}$Kavli Institute for Astronomy and Astrophysics, Peking University, Yi He Yuan Road 5, Hai Dian District, China\\
$^{9}$Institute of Theoretical Physics, Faculty of Mathematics and Physics, Charles University in Prague, Czech Republic\\
$^{10}$Warsaw University Observatory, Al Ujazdowskie 4, 00-478 Warsaw, Poland\\
$^{11}$The American Association of Variable Star Observers, 49 Bay State Road, Cambridge, MA 02138, USA\\
$^{12}$ASC Technology Services, 433 Mendenhall Laboratory 125 South Oval Mall Columbus OH, 43210, USA\\
}

\date{Accepted XXX. Received YYY; in original form ZZZ}

\pubyear{2019}

\begin{document}
\label{firstpage}
\pagerange{\pageref{firstpage}--\pageref{lastpage}}
\maketitle

\begin{abstract}
The All-Sky Automated Survey for Supernovae (ASAS-SN) provides long baseline (${\sim}4$ yrs) light curves for sources brighter than V$\lesssim17$ mag across the whole sky. As part of our effort to characterize the variability of all the stellar sources visible in ASAS-SN, we have produced ${\sim}30.1$ million V-band light curves for sources in the southern hemisphere using the APASS DR9 catalog as our input source list. We have systematically searched these sources for variability using a pipeline based on random forest classifiers. We have identified ${\sim} 220,000$ variables, including ${\sim} 88,300$ new discoveries. In particular, we have discovered ${\sim}48,000$ red pulsating variables, ${\sim}23,000$ eclipsing binaries, ${\sim}2,200$ $\delta$-Scuti variables and ${\sim}10,200$ rotational variables. The light curves and characteristics of the variables are all available through the ASAS-SN variable stars database (\url{https://asas-sn.osu.edu/variables}). The pre-computed ASAS-SN V-band light curves for all the ${\sim}30.1$ million sources are available through the ASAS-SN photometry database (\url{https://asas-sn.osu.edu/photometry}). This effort will be extended to provide ASAS-SN light curves for sources in the northern hemisphere and for V$\lesssim17$ mag sources across the whole sky that are not included in APASS DR9. 

\end{abstract}




\section{Introduction}

Recent large scale sky surveys such as the All-Sky Automated Survey (ASAS; \citealt{2002AcA....52..397P}), the Optical Gravitational Lensing Experiment (OGLE; \citealt{2003AcA....53..291U}),the Northern Sky Variability Survey (NSVS; \citealt{2004AJ....127.2436W}), MACHO \citep{1997ApJ...486..697A}, EROS \citep{2002A&A...389..149D}, the Catalina Real-Time Transient Survey (CRTS; \citealt{2014ApJS..213....9D}), the Asteroid Terrestrial-impact Last Alert System (ATLAS; \citealt{2018PASP..130f4505T,2018arXiv180402132H}), and Gaia \citep{2018arXiv180409365G,2018arXiv180409373H,gdr2var} have revolutionized the study of stellar variability. Over time, these surveys have collectively discovered $\gtrsim 10^6$ variable stars across the whole sky.

Variable stars have been used to study astrophysics in multiple contexts. Pulsating variables, including Cepheids, RR Lyrae stars and Mira variables are used as distance indicators as they follow distinct period-luminosity relationships (e.g., \citealt{1908AnHar..60...87L,2006MNRAS.370.1979M,2018SSRv..214..113B,2008MNRAS.386..313W}, and references therein). Eclipsing binary stars are used to study stellar systems and with sufficient radial velocity followup, allow for the derivation of dynamical information and fundamental stellar parameters, including masses and radii of the stars in these systems \citep{2010A&ARv..18...67T}. The precise measurements afforded by studying eclipsing binaries allow for the test of stellar theory across the Hertzsprung-Russell diagram. Variable stars are also used to study stellar populations and Galactic structure \citep{2018MNRAS.479..211M,2018IAUS..334...57M,2014IAUS..298...40F}.

The All-Sky Automated Survey for SuperNovae (ASAS-SN, \citealt{2014ApJ...788...48S, 2017PASP..129j4502K}) monitored the visible sky to a depth of $V\lesssim17$ mag with a cadence of 2-3 days using two units in Chile and Hawaii each with 4 telescopes. ASAS-SN has recently expanded to 5 units with 20 telescopes. All the current ASAS-SN units are equipped with g-band filters and are currently monitoring the sky to a depth of $g\lesssim18.5$ mag with a cadence of $\sim1$ day. The ASAS-SN telescopes are hosted by the Las Cumbres Observatory (LCO; \citealt{2013PASP..125.1031B}) in Hawaii, Chile, Texas and South Africa. The primary focus of ASAS-SN is the detection of bright supernovae and other transients (e.g., tidal disruption events, cataclysmic variables, AGN flares, stellar flares, etc.) with minimal bias (e.g., \citealt{2014MNRAS.445.3263H,2016MNRAS.455.2918H,2017MNRAS.471.4966H,2018arXiv181108904H,2018arXiv180802890H}), but its excellent baseline and all-sky coverage allows for the characterization of stellar variability across the whole sky. 
 
In Paper I \citep{2018MNRAS.477.3145J}, we discovered ${\sim}66,000$ new variables that were flagged during the search for supernovae, most of which are located in regions that were not well-sampled by previous surveys. In Paper II \citep{2019MNRAS.486.1907J}, we homogeneously analyzed ${\sim} 412,000$ known variables from the International Variable Stars Index (VSX,\citealt{2006SASS...25...47W}), and developed a versatile random forest variability classifier utilizing the ASAS-SN V-band light curves and data from external catalogues. As data from The Transiting Exoplanet Survey Satellite (TESS; \citealt{2015JATIS...1a4003R}) became available, we have explored the synergy between the two surveys. The ASAS-SN light curves have long time baselines (${\gtrsim}4$ yr) and are sampled at a cadence of ${\sim} 1-3$ days. Thus, these light curves complement the high cadence TESS light curves that have a shorter baseline.  In Paper III \citep{2019MNRAS.485..961J}, we characterized the variability of ${\sim}1.3$ million sources within 18 deg of the Southern Ecliptic Pole towards the TESS continuous viewing zone and identified ${\sim} 11,700$ variables, including ${\sim} 7,000$ new discoveries. We also identified the most extreme heartbeat star system thus known, and characterized the system using both ASAS-SN and TESS light curves \citep{2019arXiv190100005J}. We have also explored the synergy between ASAS-SN and large scale spectroscopic surveys using data from APOGEE \citep{2015AJ....150..148H} with the discovery of the first likely non-interacting binary composed of a black hole with a field red giant \citep{2018arXiv180602751T} and the identification of 1924 APOGEE stars as periodic variables in Paper IV \citep{2019MNRAS.tmp.1644P}. During our search for variables, we have also identified numerous unusual, rare variables, including 2 very long period detached eclipsing binaries \citep{2018RNAAS...2c.181J,2018RNAAS...2c.125J} and 19 R Coronae Borealis stars \citep{2018arXiv180904075S}.

Here, we extracted the ASAS-SN light curves of ${\sim}30.1$ million sources from the AAVSO Photometric All-Sky Survey (APASS; \citealt{2015AAS...22533616H}) DR9 catalog with $V<17$ mag in the southern hemisphere ($\delta<0$ deg).  In this work, we systematically search this sample for variable sources. In Section $\S2$, we discuss the input catalogue and the data reduction procedures used to obtain the ASAS-SN light curves. Section $\S3$ discusses the random forest based variability identification and classification procedures. In Section $\S4$, we discuss the ASAS-SN catalogue of variable stars in the southern hemisphere and present a summary of our work in Section $\S5$. 

\section{Observations and Data reduction}

\begin{figure*}
	\includegraphics[width=\textwidth]{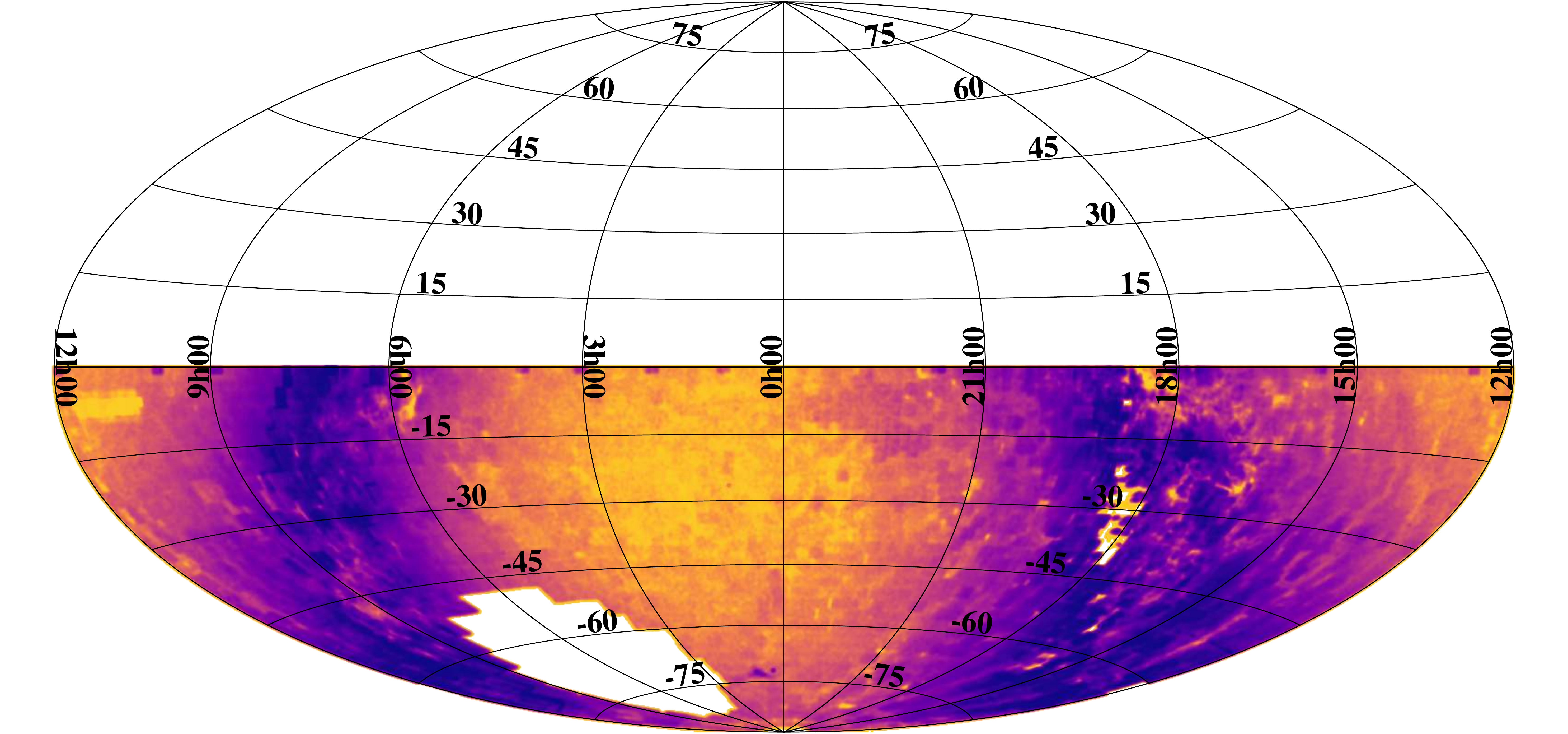}
    \caption{Sky density plot of the ${\sim}30.1$M APASS DR9 sources analyzed in this work. Sources in the gap centered at the Southern Ecliptic Pole ($\alpha=6$ h, $\delta=-66.55$ deg) were analyzed in \citet{2019MNRAS.485..961J}. The gaps near ($\alpha=18$ h, $\delta=-30$ deg) are in the input APASS DR9 catalog. }
    \label{fig:fig1}
\end{figure*}

We started with the APASS DR9 catalog as our input source catalog. We chose the APASS catalog because the APASS survey had a faint completeness limit ($V\lesssim16$) comparable to the ASAS-SN observations. We selected all the APASS sources with $V<17$ mag in the southern hemisphere ($\delta<0$ deg), excluding the ${\sim}1.3$M sources towards the Southern Ecliptic Pole which were analyzed in Paper III. This resulted in a list of ${\sim}30.1$M sources. Figure \ref{fig:fig1} illustrates the spatial distribution of these sources. 

The ASAS-SN V-band observations used in this work were made by the ``Brutus" (Haleakala, Hawaii) and ``Cassius" (CTIO, Chile) quadruple telescopes between 2013 and 2018. Each ASAS-SN V-band field is observed to a depth of $V\lesssim17$ mag. The field of view of an ASAS-SN camera is 4.5 deg$^2$, the pixel scale is 8\farcs0 and the FWHM is typically ${\sim}2$ pixels. ASAS-SN tends to saturate at ${\sim} 10-11$ mag, but we attempt to correct the light curves of saturated sources for bleed trails (see \citealt{2017PASP..129j4502K}). The V-band light curves were extracted as described in \citet{2018MNRAS.477.3145J} using image subtraction \citep{1998ApJ...503..325A,2000A&AS..144..363A} and aperture photometry on the subtracted images with a 2 pixel radius aperture. The APASS catalog was also used for calibration. We corrected the zero point offsets between the different cameras as described in \citet{2018MNRAS.477.3145J}. The photometric errors were recalculated as described in \citet{2019MNRAS.485..961J}.

While we decided to use the APASS DR9 catalog as our input source list due to its all-sky coverage, this catalog has several shortcomings \citep{2015AAS...22533616H,2019A&A...621A.144M}. While the APASS DR9 sky coverage is nearly complete, there are regions towards the Galactic plane that are missing (see Figure \ref{fig:fig1}). In addition, the DR9 catalog includes a number of duplicate entries, which appear to be caused by the merging process, where poor astrometry in a given field may cause two centroids to be included for a single source. Centroiding in crowded fields is also poor and blends cause both photometric and astrometric errors. The APASS DR9 catalog does not provide unique identifiers, thus we used the VizieR \citep{2018AJ....155...39O} \verb"recno" field as unique identifiers. To address the issue of incomplete sky coverage we will use the ATLAS All-Sky Stellar Reference Catalog \citep{2018ApJ...867..105T} in the next paper to produce light curves for the missing sources in APASS DR9.

\section{Variability Analysis}

Here we describe the procedure we used to identify and characterize variables in the source list. We describe how we cross-matched the APASS sources to external catalogues in Section $\S3.1$. In Section $\S3.2$, we describe the procedure we took to identify candidate variable sources using a random forest classifier. In Section $\S3.3$, we discuss the classification of the candidate variables into the various standard classes of variable stars using the V2 random forest classifier model from \citet{2019MNRAS.486.1907J}, in Section $\S3.4$, we discuss our attempts to mitigate the effects of blending on the list of candidate variables and in Section $\S3.5$, we discuss the quality checks that we used to improve the final variables catalog.

\subsection{Cross-matches to external catalogs}

We cross-match the APASS DR9 sources with Gaia DR2 \citep{2018arXiv180409365G} using a matching radius of 5\farcs0. The sources were also cross-matched to the Gaia DR2 probabilistic distance estimates from \citet{2018AJ....156...58B}. Even though we used a liberal matching radius, ${\sim}84\%$ (${\sim}94\%$) of the sources have a cross-match in Gaia DR2 within 2\farcs0 (3\farcs0).
These sources were also cross-matched with the 2MASS \citep{2006AJ....131.1163S} and AllWISE \citep{2013yCat.2328....0C,2010AJ....140.1868W} catalogues using a matching radius of 10\farcs0. The cross-matches to these catalogues provide useful information that are later used in the identification and classification of variable stars. We used \verb"TOPCAT" \citep{2005ASPC..347...29T} to cross-match the APASS sources with these external catalogues. 

Sources in the Small Magellanic Cloud (SMC) are also included in our input source list. We used association information from Gaia DR2 \citep{2018A&A...616A..12G} to identify ${\sim}1,600$ sources from our source list that are SMC members. For sources in the SMC, we use a distance of $d=62.1$ kpc \citep{2014ApJ...780...59G} in our variability classifier instead of the distance estimate from \citet{2018AJ....156...58B}. The LMC was covered in Paper III.

\subsection{Random Forest Variable Identification}

In Paper III we used several methods, including linear cuts on Lomb-Scargle periodogram statistics, light curve features and external photometry to identify variable sources. Here, we take a different approach by training and apply a random forest classifier to distinguish candidate variables from constant sources. We built a variability classifier based on a random forest (RF) model using \verb"scikit-learn"  \citep{2012arXiv1201.0490P,breiman}. A random forest classifier is an ensemble of decision trees whose output is the mean prediction of the individual decision trees \citep{breiman}. The set of variable sources used to train this classifier consisted of ${\sim} 302,000$ variables from Papers II and III with definite classifications. Variables with uncertain classifications, including `VAR' and `ROT:', were not included in this list as they reduced the accuracy of the final random forest classification model. The set of constant sources in the training list consisted of ${\sim} 600,000$ sources randomly selected from the list of constant sources in Paper III.

The goal was to provide classifications into two broad groups: CONST (constant stars) and VAR (potential variables). The potential variables will be analyzed in further detail so it is more important not to lose real variables than to accidentally include non-variables. These broad classes were selected to reduce the complexity of the classifier, and to provide an accurate initial separation prior to reclassifying the variable sources with the random forest variable type classifier from Paper II. To generate periodicity statistics, we used the \verb"astropy" implementation of the Generalized Lomb-Scargle (GLS, \citealt{2009A&A...496..577Z,1982ApJ...263..835S}) periodogram to search for periodicity over the range $0.05\leq P \leq1000$ days in all ${\sim}30.1$M light curves. The GLS periodogram is an extension of the standard Lomb-Scargle periodogram that uses a frequency dependent light curve mean, $$y_{\rm model}(t; f) = y_0(f) + A_f \sin \big( 2\pi f(t -\phi_f ) \big),$$ (eq. 41 from \citealt{2018ApJS..236...16V}). This floating-mean model is more robust if there are gaps in the phased data than the standard Lomb-Scargle periodogram \citep{2018ApJS..236...16V}. We utilize the best GLS period, false alarm probability (FAP) and the power of the best GLS period as features. The best GLS period is defined to be the period with the largest power which is essentially a measure of the SNR of the periodogram peak. The false alarm probability is the probability that a light curve with no signal would lead to a GLS peak of a similar magnitude \citep{2018ApJS..236...16V}.

We further characterize the periodicity of the light curves using the Lafler-Kinmann string length statistic \citep{1965ApJS...11..216L,2002A&A...386..763C}. We use the definition \begin{equation}
    T(\phi|P)=\frac{\sum_{i=1}^{\rm N} (m_{i+1}-m_i)^2}{\sum_{i=1}^{\rm N} (m_{i}-\overline m)^2}\times \frac{(N-1)}{2N}
	\label{eq:tt}
\end{equation} from \citet{2002A&A...386..763C}, where the $m_i$ are the magnitudes sorted by phase and $\overline m$ is the mean magnitude. We also calculate the statistic $T(t)$ after sorting the light curve by time instead of phase. The complete list of 20 features and their importances to the trained random forest model is summarized in Table \ref{tab:features}. Feature importances are calculated as Gini importances using the mean decrease impurity algorithm \citep{2012arXiv1201.0490P}.

The overall results of the random forest model are evaluated based on the \begin{equation}
    \rm precision=\frac{\alpha}{\alpha + \beta} \,,
	\label{eq:prec}
\end{equation} \begin{equation}
    \rm recall=\frac{\alpha}{\alpha + \gamma}\,,
	\label{eq:rec}
\end{equation} and the harmonic mean of the two, \begin{equation}
    F_1=2 \bigg( \frac{\rm precision\times \rm recall}{\rm precision+\rm recall}\,\bigg),
	\label{eq:f1}  \end{equation} where $\alpha$, $\beta$ and $\gamma$ are the number of true positives, false positives and false negatives in a given class respectively. These quantities are evaluated for both the constant (CONST) and variable (VAR) sources (Table \ref{tab:perf}).
	
\begin{figure}
	\includegraphics[width=0.5\textwidth]{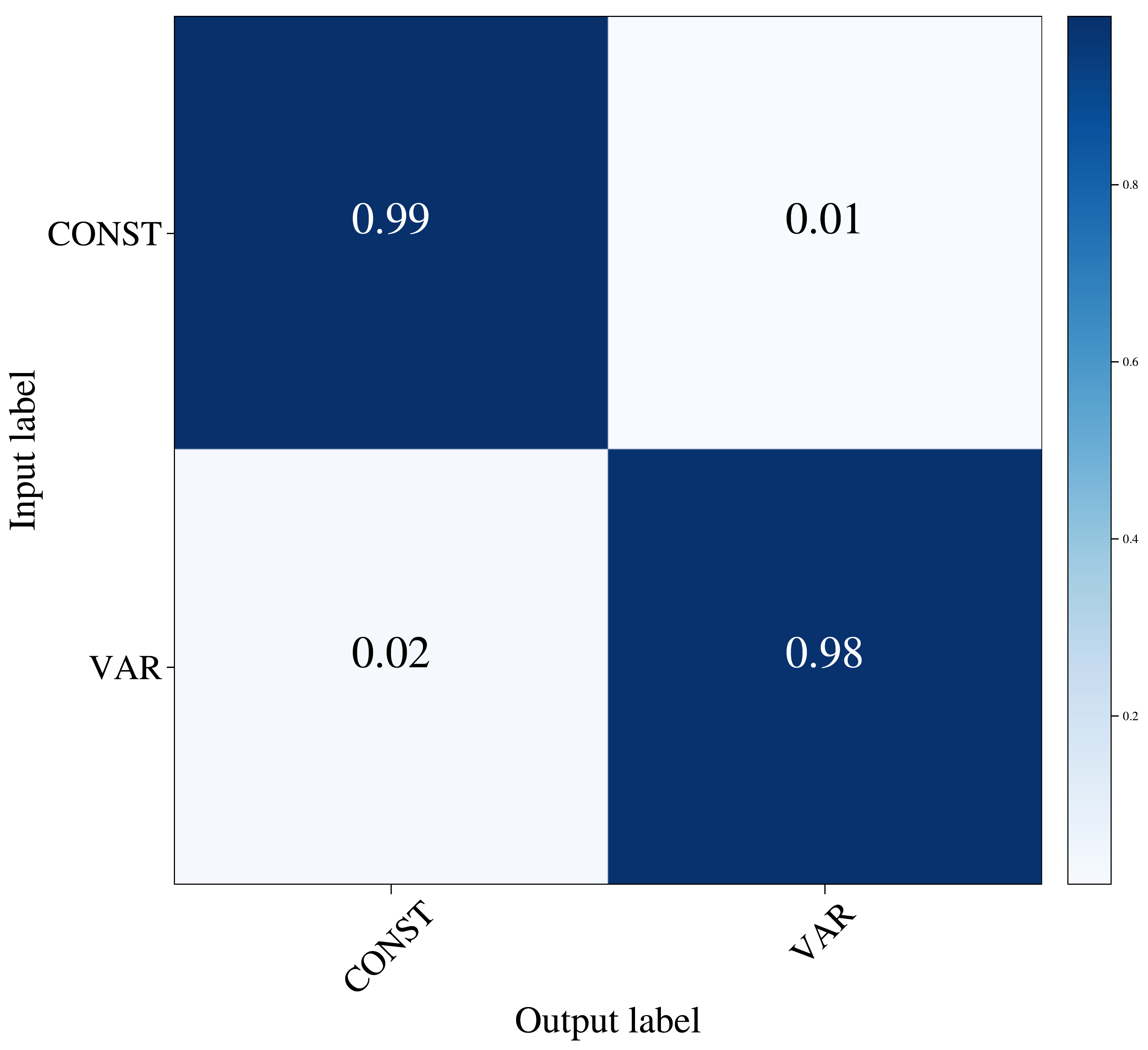}
    \caption{The normalized confusion matrix derived from the final version of the trained random forest classifier. The y-axis corresponds to the `input' classification, while the x-axis is the `output' prediction obtained from the trained random forest model.}
    \label{fig:fig2}
\end{figure}

The parameters of the random forest model were optimized using cross-validation to maximize the overall $F_1$ score of the classifier. The number of decision trees in the forest was initialized to \verb"n_estimators=1000". We also limited the maximum depth of the decision trees to \verb"max_depth=16" in order to mitigate over-fitting, set the number of samples needed to split a node as \verb"min_samples_split=10" and set the number of samples at a leaf node as \verb"min_samples_leaf=5". To further minimize over-fitting, we also assigned weights to each class with \verb"class_weight=`balanced_subsample'". For any given source, the RF classifier assigns classification probabilities $\rm Prob(Const)$ and $\rm Prob(Var)=1-Prob(Const)$. The output classification of the RF classifier is the class with the highest probability. We split the training sample for training ($80\%$) and testing ($20\%$) in order to evaluate the performance of the RF classifier. The confusion matrix for the trained RF model is shown in Figure \ref{fig:fig2}. The greatest confusion (2\%) arises from input variable sources that are subsequently classified as constant stars. The performance of the classifier is summarized in Table \ref{tab:perf}. The overall $F_1$ score for the classifier is 98.5\%. 

We further investigate the performance variation of the RF classifier with magnitude by binning the test sample by the median V-band magnitude (Table \ref{tab:perfmag}). The $F_1$ score for the classifier is lowest ($92.9\%$) for sources with $V\leq11$ mag where 15\% of the input constant sources are incorrectly classified as variable sources. This is likely due to saturation artifacts that arise at these bright magnitudes. The $F_1$ score varies by $<1\%$ between $11<V\leq17$ mag and the confusion between constant and variables sources is also minimal. Most of the sources ($96.3\%$) in the test sample have median magnitudes that fall within this range. For sources at the ASAS-SN V-band faint limit of $V\geq17$ mag, the $F_1$ score drops to $95.5\%$ and the confusion between classes again rises. This is not surprising as at these magnitudes the light curves are dominated by noise.

\begin{table*}
	\centering
    \setlength\tabcolsep{2pt}
	\caption{Variability features and their importances for variable star identification}
	\label{tab:features}
	\begin{tabular}{||p{2cm}| p{10cm}| c| p{4cm}||} 
		\hline
		Feature & Description & Importance & Reference \\
		\hline
        LS\_Per & Best Lomb-Scargle period & 1\% &  -  \\
        LS\_Pow & Power corresponding to the best Lomb-Scargle period & 24\%  &  -  \\        
        log(LS\_FAP) & Base 10 logarithm of the False Alarm Probability corresponding to the best Lomb-Scargle period & 11\%   &  -  \\  
        $\rm T(t)$ & Lafler-Kinman String Length statistic of the light curve sorted by time & 4\% & \citet{2019MNRAS.486.1907J}  \\          
        $\rm T(\phi|P)$ & Lafler-Kinman String Length statistic of the light curve sorted by phase & 18\% & \citet{2019MNRAS.486.1907J}  \\          
        $\delta$ & Normalized difference between $\rm T(t)$ and $\rm T(\phi|P)$ & 5\% &\citet{2019MNRAS.485..961J} \\      
        $\rm Skew$ & Skewness of the magnitude distribution & 2\%  &-\\ 
        $\rm Kurt$ & Kurtosis of the magnitude distribution & 2\%  &-\\  
        $\rm Median$ & Median of the magnitude distribution & 1\%  &-\\ 
        $\sigma$ & Standard deviation of the light curve  & 2\% & -  \\            
        IQR & Difference between the 75$^{\rm th}$ and 25$^{\rm th}$ percentiles in magnitude & 2\%   &-\\   
        $A_{\rm HL}$ & Ratio of magnitudes brighter or fainter than the average & 2\%  & \citet{upsilon} \\
        $\rm MAD$ & Median absolute deviation of the light curve  & 2\% & -  \\    
        $1/\eta$ & Inverse of the $\eta$ (Von Neumann index) value for the light curve  & 3\%  & \citet{vn}  \\   
        $J-K_s$ & 2MASS $J-K_s$ color  & 12\%  &\citet{2006AJ....131.1163S} \\
        $H-K_s$ & 2MASS $H-K_s$ color  & 9\% &\citet{2006AJ....131.1163S} \\          
		\hline
	\end{tabular}
\end{table*} 

\begin{table}
	\centering
	\caption{Overall performance of the ASAS-SN random forest source classifier}
	\label{tab:perf}
	\begin{tabular}{lrrrrrrr}
		\hline
		Class & Precision & Recall & $F_1$ score & Sources\\
		\hline
        CONST & 99$\%$ & 99$\%$ & 99$\%$ & 600,000\\
		VAR & 98$\%$ & 98$\%$ & 98$\%$ & 302,021 \\
		\hline
	\end{tabular}

\end{table}

\begin{table*}
	\centering
	\caption{Performance of the ASAS-SN random forest source classifier with magnitude}
	\label{tab:perfmag}
	\begin{tabular}{lrrrrrrr}
		\hline
		Median magnitude & $F_1$ score & Constant Star False Positive Rate & Variable Star False Positive Rate & Sources (\%)\\
		\hline
        $V\leq11$ mag & 92.9$\%$ & 15$\%$ & 2$\%$ & 3.1\%\\
		$11<V\leq13$ mag & 98.5$\%$ & 1$\%$ & 2$\%$ & 11.5\%\\
		$13<V\leq15$ mag & 98.9$\%$ & 1$\%$ & 1$\%$ & 37.5\%\\		
		$15<V\leq17$ mag & 98.4$\%$ & 1$\%$ & 3$\%$ & 47.3\%\\
		$V\geq17$ mag & 95.5$\%$ & 5$\%$ & 4$\%$ & 0.8\%\\			
		\hline
	\end{tabular}

\end{table*}

\begin{figure*}
	\includegraphics[width=\textwidth]{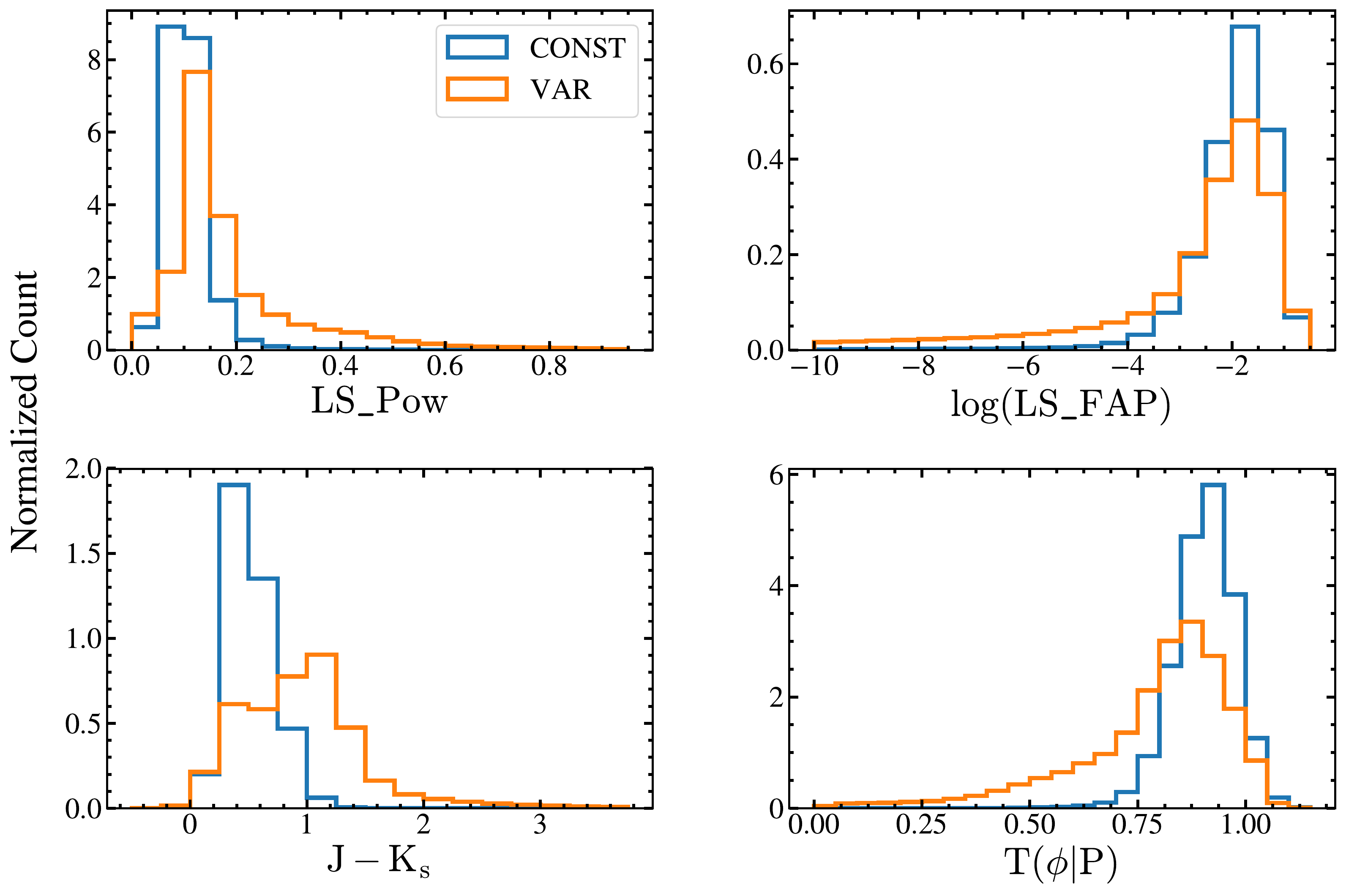}
\caption{Distribution of the sources classified as CONST and VAR in the LS\_Pow, $\rm T(\phi|P)$, $J-K_s$ and log(LS\_FAP) features. Sources with log(LS\_FAP)$<-10$ are not shown for clarity.}    
    \label{fig:fig3}
\end{figure*}

We applied the trained random forest classifier to the entire sample of ${\sim}30.1$M sources and identified 3,553,235 candidate variables. The distinction between the constant sources and the candidate variables is illustrated in Figure \ref{fig:fig3} through the distributions of the four features with the largest importance: LS\_Pow, $\rm T(\phi|P)$, $J-K_s$ and log(LS\_FAP). We find that many candidate variable sources are strongly periodic with high values of LS\_Pow and smaller values of log(LS\_FAP) and $\rm T(\phi|P)$. In addition, the distribution of the 2MASS color $J-K_s$ differs significantly between constant and variable sources. Variable sources are skewed towards redder near-infrared colors with $J-K_s>1$ mag while constant sources largely peak around $J-K_s\sim0.5$ mag. Cooler, evolved stars are more likely to be variable, so this is not unexpected. Nevertheless, the distributions of constant and variable sources overlap significantly in the feature spaces illustrated in Figure \ref{fig:fig3}. It is crucial to note that while we illustrated these distributions linearly, the RF classifier is inherently non-linear and relies on a complex ensemble of decision trees to predict the class of any given source. In this sense, even the features with the least importance do matter for the overall success of the classifications.

\subsection{Variability Classification}

Once candidate variables are identified, we aimed at classifying these sources into the various standard classes of variable stars. We use the variability classifier implemented in \citet{2019MNRAS.486.1907J}, which consists of a random forest classifier plus several refinement steps. Given the large number of candidates, we changed our variability classification strategy as follows. 
\begin{itemize}
  \item Initially, we classified all the candidate variables using just the GLS periods derived in $\S 3.2$.
  \item Following this, we derive periods for a limited set of sources (see below) using the \verb"astrobase" implementation \citep{astrob} of the Box Least Squares (BLS, \citealt{2002A&A...391..369K}) periodogram to improve the completeness for eclipsing binaries whose periodicity cannot be easily identified with GLS.
\end{itemize}
We also run the variability classifier twice, once using the best period (GLS or BLS) and once using twice the best period. The final classification is the one which yields the greatest classification probability. This step greatly improves the separation of EW type eclipsing binaries from RRC variables, and also improves upon the efficiency of the automated period doubling algorithm that was used for eclipsing binaries in Paper II.

To identify possible eclipsing binaries, we selected ${\sim}576,000$ candidate variables with $A_{\rm HL}>2$ (Table \ref{tab:features}) from our original list. $A_{\rm HL}$ is the ratio of magnitudes brighter or fainter than the average magnitude \citep{upsilon}. Since eclipses usually only span limited phase ranges, $A_{\rm HL}$ tends to be higher for the light curves of eclipsing binaries compared to other variable types. We searched for periods over the range $0.05\leq P \leq1000$ days and the BLS periodogram was initialized with 200 phase bins and a minimum (maximum) transit duration of $0.1$ $(0.3)$ in phase. BLS periods were only selected if the BLS power was $<0.3$. We identified ${\sim}3,500$ eclipsing binaries through this process.

\subsection{Blending Corrections}
Blending towards crowded regions (e.g., the Galactic disk) is problematic owing to the large pixel scale (8\farcs0) and the FWHM (${\sim}$16\farcs0) of the ASAS-SN images. The APASS data has a significantly smaller pixel scale (2\farcs6), so we can have multiple APASS sources inside a single ASAS-SN resolution element. To minimize the number of false positives in our catalogue of variables due to blending, we identify and correct blended variable groups in our catalog. However, we have not attempted to correct for the contaminating light in the photometry of the blended sources. 

We find that ${\sim}1.1$M of the ${\sim}3.6$M candidate variables had at least one neighbor within 30\farcs0 from their positions. For these sources, we compute the flux variability amplitudes using a non-parametric random forest regression model \citep{2019MNRAS.486.1907J}. We then identify groups of blended variables based on the cross-matching. For each blended variable group, we select the source with the largest flux variability as the `true' variable, and discard the remaining overlapping sources in the blended group from the final list. After removing these overlapping, blended sources from the list of candidate variables, we were left with ${\sim}3$M sources.

\subsection{Quality Checks}

At this stage, visual review of a random set of light curves suggested that quality checks must be implemented to distinguish \textit{true} variability signals from variability due to bad photometry, and other survey specific issues (e.g., shutter failures, etc.). In Paper III, given the significantly shorter list of candidate variables, this was accomplished through simple visual review of the light curves. In this work, given the shear number of sources, visual review is not a feasible option. Thus, we choose to implement various criteria in lieu of visual review, to distinguish the true variables from the `noise'.

We first restrict the list to sources with $\rm V_{mean}>10$ mag, $A>0.05$ mag and $\rm T(t)<0.9$. We implemented the cut in the ASAS-SN V-band magnitude to minimize noise due to saturation artifacts. We also calculate the ratio between the amplitude estimated by random forest regression ($A$) to the interquartile range IQR (Table \ref{tab:features}) of the light curve,\begin{equation}
    \alpha= A/\rm IQR \,, 
	\label{eq:alp}
\end{equation} and the absolute, reddening-free Wesenheit magnitudes \citep{1982ApJ...253..575M,2018arXiv180803659L} \begin{equation}
    W_{RP}=M_{\rm G_{RP}}-1.3(G_{BP}-G_{RP}) \,, 
	\label{eq:wrp}
\end{equation} 
and
\begin{equation}
    W_{JK}=M_{\rm K_s}-0.686(J-K_s) \,,
	\label{eq:wk}
\end{equation}for each source, where the $G_{BP}$ and $G_{RP}$ magnitudes are from Gaia DR2 \citep{2018arXiv180409365G} and the $J$ and $K_s$ magnitudes are from 2MASS \citep{2006AJ....131.1163S}. The Wesenheit magnitudes are used in the pipeline from paper II to refine variable type classifications. The quantity $\alpha$ can be used to identify light curves with significant outliers as we expect $\alpha \approx 2$ for most sources.

The criteria used in lieu of visual review are summarized in Table \ref{tab:refsum}. We note that these criteria are applied in addition to the refinement criteria in Paper II. These are not replacements but additional quality checks intended to improve the purity of our catalog. We derived these criteria through visual inspection in order to minimize false positives in the different variable groups. In addition to the criteria summarized in Table \ref{tab:refsum}, we further scrutinize sources with periods that are close to aliases of a sidereal day (e.g., $P\approx1$ d, $P\approx2$ d, $P\approx30$ d, etc.). This is accomplished by tightening the criteria on $\rm T(\phi|P)$, log(LS\_FAP), LS\_Pow and $\delta$. This process slightly reduces the completeness of our catalog at these periods, but greatly reduces the number of false positives. In addition, we removed QSO contaminants in this list by cross-matching our list of variables to the \citet{2019RAA....19...29L} catalog of known QSOs using a matching radius of 5\farcs0. We identified 336 cross-matches, out of which 325 were classified as YSO variables in our pipeline. At this point we had ${\sim}247,200$ sources nominally classified as variable stars.

We inspected 5000 randomly selected sources classified as non-variable and the same number classified as variable. This was partly just a sanity check but also driven by the concern that the large size of our initial list (${\sim}10\%$ of the sources) suggested that our false positive rates had to be higher than suggested by Table \ref{tab:perf}. Among the non-variable sources, we identified only 3 ($\sim 0.06\%$) that might be low level variables, which suggests that we are missing few variables that can be detected in this data. For the variable stars, we found significant numbers of false positives in the following variable classes: GCAS (${\sim}50\%$), L (${\sim}25\%$), VAR (${\sim}45\%$) and YSO (${\sim}13\%$). The implied false positive rate
for the variable sources was ${\sim}5.9\%$ at this point.

 Light curves that are contaminated by systematics tend to be classified as irregular or generic variables as they are inherently aperiodic in nature. Thus, we decided to review all ${\sim}32,800$ sources that were classified as L, VAR, GCAS, or YSO to improve the purity of our catalog. Initial results suggested that L variables with $T(t)>0.65$ were dominated by noise, so we rejected ${\sim}14,300$ such sources without further visual review. We visually reviewed the remaining ${\sim}18,500$ sources, and rejected ${\sim}12,600$ sources (${\sim}68\%$) and only retained ${\sim}5,900$ of these sources in the final catalog. When we carried out a new inspection of 5000 randomly selected variables, the false positive rates were now EA (${\sim}1.4\%$), L (${\sim} 0.6\%$), SR (${\sim} 2.6\%$), and VAR (${\sim} 0.9\%$). This implies an overall false positive rate for the final catalog of variable sources of $\sim 1.3\%$.

After these criteria are applied, we end up with a list of ${\sim}220,000$ variables. This means that our initial candidate list had a false positive rate of ${\sim}93\%$. The larger than expected false positive rate is partly due to a biased training set in the source classifier. The training set of constant sources was derived from a region of the sky away from the Galactic plane. The increased crowding and blending towards the Galactic plane will systematically affect constant stars at low latitudes and introduce spurious variability signals into their light curves. Our classifier will identify these constant sources as candidate variables. In addition to this, sources in the vicinity of bright, saturated stars in our data are likely to have spurious variability signals in their image subtraction light curves due to the corrections made for bleed trails (see \citealt{2017PASP..129j4502K}). This effect is again exacerbated towards the Galactic plane.

\begin{table*}
	\centering
	\caption{Summary of the variability refinement criteria for each variable class.}
	\label{tab:refsum}
\begin{tabular}{||p{5cm}| p{12cm}||}
		\hline
		Class & Summarized refinement Criteria \\\\
				\hline
$\delta$ Scuti (HADS, DSCT)  & \\		
   & $\rm Skew<0.15$, LS\_Pow>0.25, log(LS\_FAP)$<-7$, $A<0.5$ mag, $\rm T(\phi|P)<0.5$,  $-1<W_{JK}<3$ mag  \\
		\hline
		
RR Lyrae (RRAB, RRC, RRD)   &  \\		
   & RRAB and log(LS\_FAP)$<-10$, LS\_Pow>0.2, $A>0.08$ mag, $\rm T(\phi|P)<0.6$, $\rm Skew<0.15$, $\delta<-0.25$  \\  
   & RRC/RRD and log(LS\_FAP)$<-10$, LS\_Pow>0.2, $A>0.08$ mag, $\rm T(\phi|P)<0.6$, $\rm Skew<0$, $\delta<-0.25$  \\   
   \hline

Cepheids (DCEP, DCEPS, CWA, CWB, RVA)    &  \\	
   & $\rm Skew<1$, log(LS\_FAP)$<-10$, LS\_Pow>0.3, $A<2$ mag, $\rm T(\phi|P)<0.6$, $\delta<-0.25$ \\ 
   \hline
   
Rotational Variables (ROT) & \\		
   &  Period$>0.6$ d and log(LS\_FAP)$<-5$, LS\_Pow>0.2, $A>0.08$ mag, $\rm T(\phi|P)<0.6$, $\delta<0$\\
   &   Period$\leq0.6$ d and $W_{JK}>2.5$ mag, $\rm Prob>0.9$\\   
		\hline   
Eclipsing Binaries (EA, EB, EW)  & \\		
   & EA (GLS) and $\alpha<100$, $\rm T(\phi|P)<0.6$, $A>0.08$ mag  \\
   & EB (GLS) and log(LS\_FAP)$<-7$, LS\_Pow>0.2, $A>0.08$ mag, $\rm T(\phi|P)<0.6$  \\  
   & EW (GLS) and log(LS\_FAP)$<-7$, LS\_Pow>0.2, $A>0.08$ mag, $\rm T(\phi|P)<0.6$, $\rm Skew>0$  \\ 
   & \\
   & EA (BLS) and $\alpha<100$, $\rm T(\phi|P)<0.45$, $\rm Prob>0.8$ \\
   \hline  
   
Semiregular and Irregular Variables (SR, L)  &  \\		
   & $\alpha<5$, $\rm V_{mean}>11$ mag, $A>0.08$ mag  \\
   &  Period$>100$ d and log(LS\_FAP)$<-3$, $J-K_s>1.1$, $A>0.1$ mag, $\rm T(t)<0.7$\\
   &  $10\leq$Period$\leq100$ d and log(LS\_FAP)$<-8$, $A>0.08$ mag\\   
   \hline  
Mira Variables (M, M:)   & \\		
   &  log(LS\_FAP)$<-3$, LS\_Pow>0.5 , $\rm T(\phi|P)<0.5$ \\
   \hline    
Young Stellar Objects (YSO) & \\		
   &  Period$<100$ d and $\alpha<5$, log(LS\_FAP)$<-10$, LS\_Pow>0.25, $\rm T(\phi|P)<0.6$\\ 
		\hline  
Outbursting Be stars (GCAS, GCAS:) & \\		
   &  $\alpha<5$, $\rm V_{mean}>11$ mag, $J-K_s<1.1$, $0.25<A<1$ mag, $\rm T(t)<0.5$\\
		\hline  		
Generic Variables (VAR) & \\		
   &  $\alpha<5$, $0.1<A<2$ mag, $W_{JK}>-4$ mag, $\rm V_{mean}>11$ mag, $\rm T(\phi|P)<0.5$ OR $\rm T(t)<0.5$\\
		\hline 		
\end{tabular}
\end{table*}

\section{Results}

The complete catalog of ${\sim}220,000$ variables and their light curves are available at the ASAS-SN Variable Stars Database (\href{https://asas-sn.osu.edu/variables}{https://asas-sn.osu.edu/variables}) along with the V-band light curves for each source. Most of the known variables identified in this work were already added to the Variable Stars Database in Paper II. We have overhauled the web interface for the ASAS-SN Variable Stars Database to include interactive light curve plotting and photometry from Gaia DR2, APASS DR9, 2MASS and ALLWISE. Table \ref{tab:var} lists the number of sources of each variability type in the catalog. 

In order to identify known variable stars, we matched our list of variables to the VSX \citep{2006SASS...25...47W} catalog, with a matching radius of 16\farcs0. The variables discovered by the All-Sky Automated Survey (ASAS; \citealt{2002AcA....52..397P}) and the Catalina Real-Time Transient Survey (CRTS; \citealt{2014ApJS..213....9D}) have already been incorporated into the VSX database. Numerous other studies (e.g., \citealt{2012MNRAS.427.3374M,2015A&A...574A..15F} and references therein) have also searched for variable stars on a smaller scale (e.g., globular clusters). While some of these results have been included in the VSX catalog, we note that the inclusion of these studies in the VSX catalog is likely to be incomplete. We also match our variables to the catalog of variable stars discovered by ASAS-SN \citep{2018MNRAS.477.3145J}, the catalogs of variable stars in the Magellanic clouds and the Galactic bulge from the Optical Gravitational Lensing Experiment (OGLE; \citealt{2003AcA....53..291U,2016AcA....66..421P,2016AcA....66..405S,2018AcA....68..315U}, and references therein), the catalog of variables from the Asteroid Terrestrial-impact Last Alert System (ATLAS; \citealt{2018PASP..130f4505T,2018arXiv180402132H}), the Gaia DR2 catalog of variables \citep{2018arXiv180409365G,2018arXiv180409373H,gdr2var}, the catalog of variables identified by KELT \citep{2018AJ....155...39O}, the catalog of WISE variables \citep{2018ApJS..237...28C} and the variables from MACHO \citep{1997ApJ...486..697A}. Of the ${\sim}220,000$ variables identified in this work, ${\sim}131,900$ were previously discovered by other surveys, and ${\sim}88,300$ are new discoveries, as also listed in Table \ref{tab:var}.

It is evident that previous surveys, including our discoveries from paper I, successfully discovered sources that have large amplitudes or are strongly periodic. Most (${\sim}54\%$) of our new discoveries are red, pulsating variables. We also discover a large number of binaries and rotational variables, amounting to ${\sim}26\%$ and ${\sim}12\%$ of the newly discovered variable sources, respectively. It is also noteworthy that we discover many more $\delta$ Scuti sources than previously known. These variables are particularly interesting as they pulsate at high frequencies (P$<0.3$ d) and are located towards the lower end of the instability strip \citep{1979PASP...91....5B}. $\delta$ Scuti variables are also known to follow a period-luminosity relationship \citep{1990A&AS...83...51L}.

\begin{table*}
	\centering
	\caption{Variables by type}
	\label{tab:var}
\begin{tabular}{llrr}
		\hline
		VSX Type & Description & Known & New \\
		\hline
CWA   & W Virginis type variables with $P>8$ d & 205 & 51\\
CWB   & W Virginis type variables with $P<8$ d & 225 & 45\\
DCEP  & Fundamental mode Classical Cepheids& 645 & 16\\
DCEPS & First overtone Cepheids & 171 & 9\\
DSCT  & $\delta$ Scuti variables & 744 & 1354\\
EA    & Detached Algol-type binaries & 11413 & 9948\\
EB    & $\beta$ Lyrae-type binaries & 8574 & 4115\\
EW    & W Ursae Majoris type binaries & 27926 & 8887\\
HADS  & High amplitude $\delta$ Scuti variables & 1338 & 849\\
M    & Mira Variables & 3472 & 38\\
ROT   & Rotational variables & 7200 & 10236\\
RRAB  & RR Lyrae variables (Type ab) & 12936 & 294\\
RRC   & First Overtone RR Lyrae variables &2655 & 1015\\
RRD   & Double Mode RR Lyrae variables & 233& 12\\
RVA   & RV Tauri variables (Subtype A) & 32 & 0\\
SR    & Semi-regular variables & 44198 & 45556\\
\hline
L     & Irregular variables & 4979 & 2528\\
GCAS  & $\gamma$ Cassiopeiae variables & 20 & 8\\
YSO   & Young stellar objects & 1949 & 980\\
\hline
GCAS:  & Uncertain $\gamma$ Cassiopeiae variables & 20 & 6\\
VAR  & Generic variables & 2789 & 2343\\
\hline
\end{tabular}
\end{table*}

\begin{figure*}
	\includegraphics[width=\textwidth]{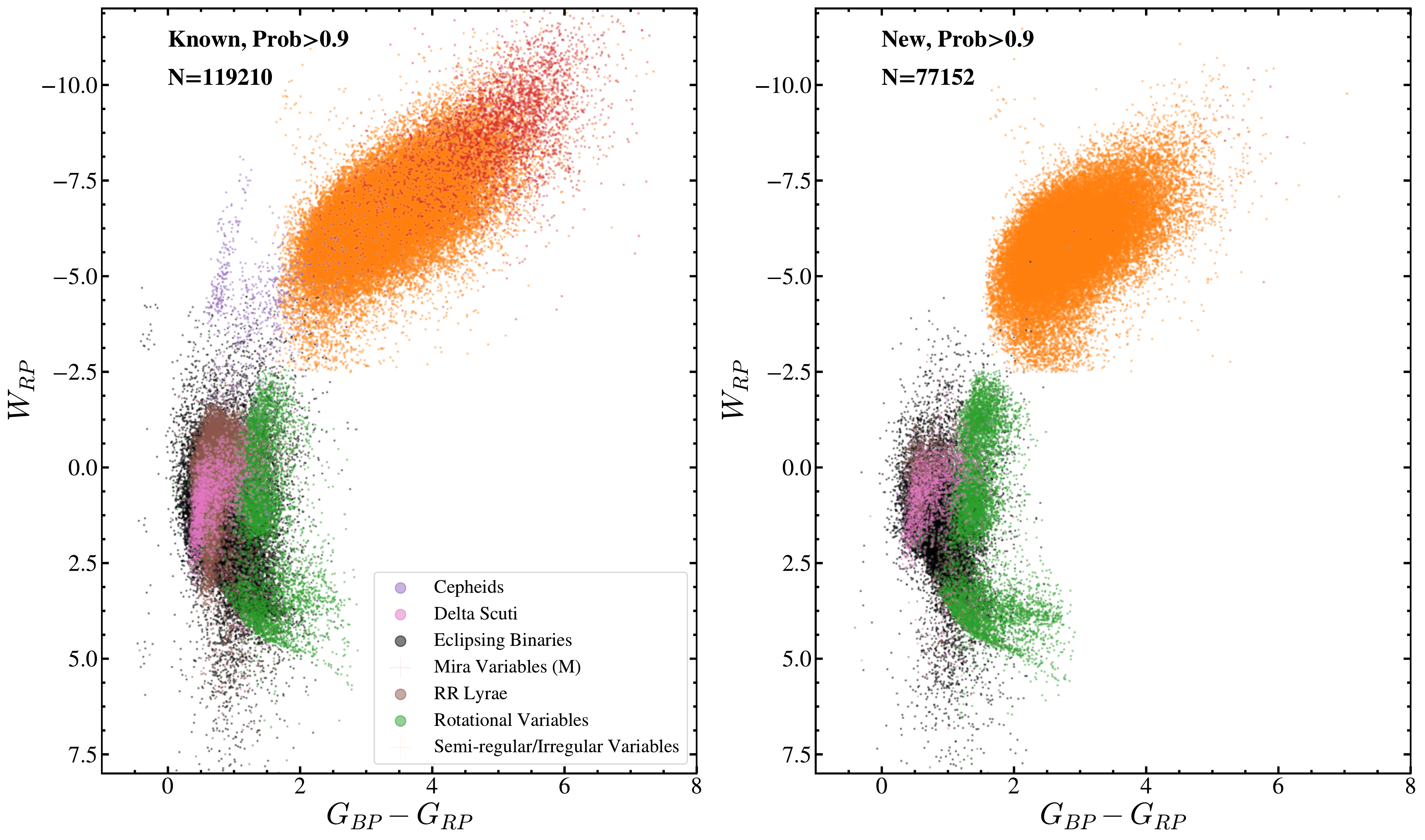}
    \caption{The Wesenheit $W_{RP}$ vs. $G_{BP}-G_{RP}$ color-magnitude diagram for the variables with $\rm Prob>0.9$, that have already been discovered (left), and the new discoveries (right).}
    \label{fig:fig4}
\end{figure*}

\begin{figure*}
	\includegraphics[width=\textwidth]{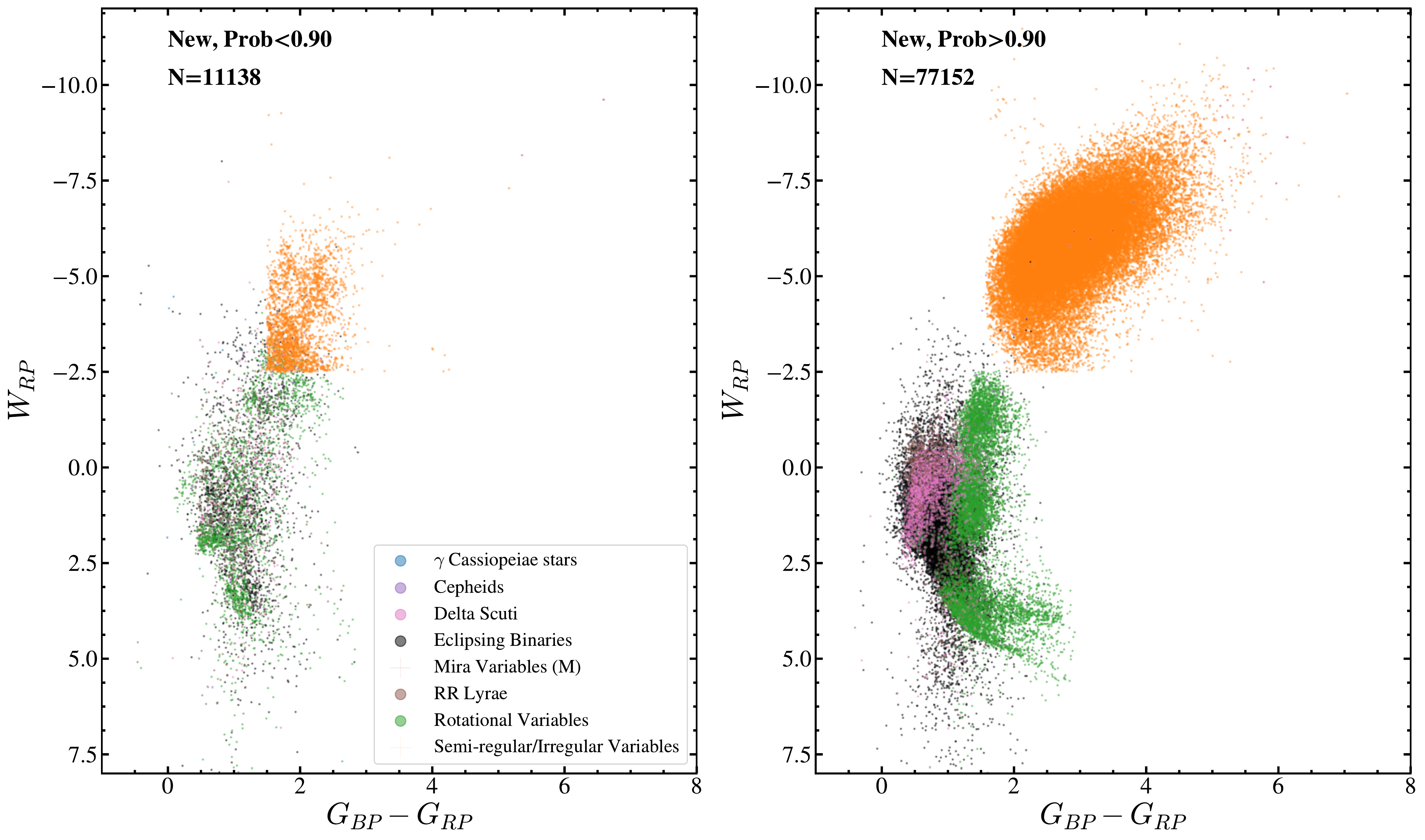}
    \caption{The Wesenheit $W_{RP}$ vs. $G_{BP}-G_{RP}$ color-magnitude diagram for the newly discovered variables with $\rm Prob<0.9$ (left), and $\rm Prob>0.9$ (right). }
    \label{fig:fig5}
\end{figure*}

The Wesenheit $W_{RP}$ vs. $G_{BP}-G_{RP}$ color-magnitude diagram for all the variables with excellent variable type classification probabilities ($\rm Prob>0.9$) is shown in Figure \ref{fig:fig4}. Generic and uncertain variable types are not shown. We have sorted the variables into groups to highlight the different classes of variable sources. A similar Wesenheit $W_{RP}$ vs. $G_{BP}-G_{RP}$ color-magnitude diagram for all the newly discovered variables, separated by probability, is shown in Figure \ref{fig:fig5}. The sharp cutoffs seen in the sample of semi-regular variables with $\rm Prob<0.9$ are inherited from the variable type refinements from paper II. Most variables with $\rm Prob<0.9$ are located in similar areas of the CMD as the variables with $\rm Prob>0.9$. However, we note two interesting clusters of these low-probability variables at ($G_{BP}-G_{RP}$, $W_{RP}$) $\sim$ $(2.5,-4.5)$ and $(0.75,1.8)$ corresponding to semi-regular and rotational variables respectively. We plan to further investigate these clusters after the variable sources in the northern hemisphere have been incorporated into our catalog. 

\begin{figure*}
	\includegraphics[width=\textwidth]{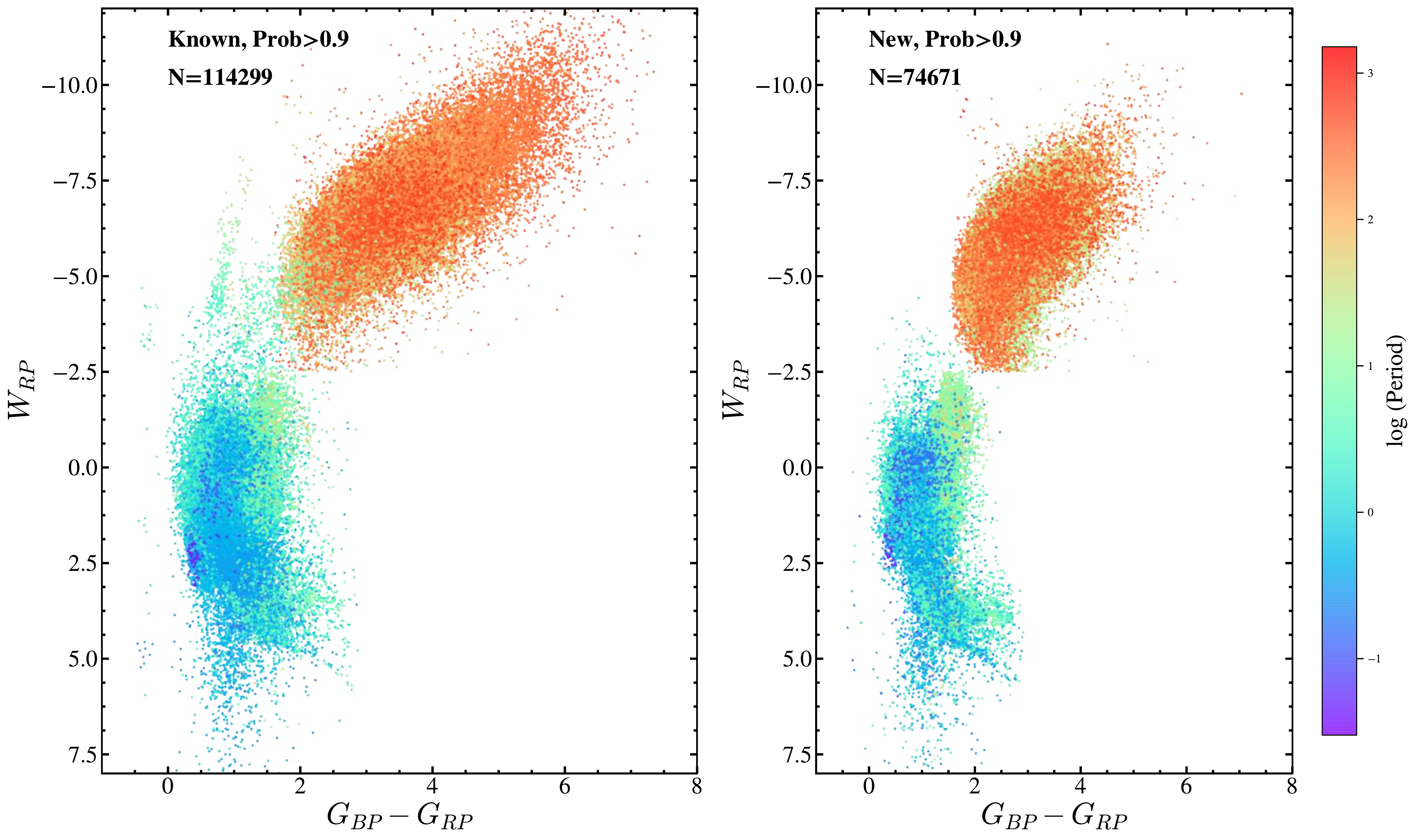}
    \caption{The Wesenheit $W_{RP}$ vs. $G_{BP}-G_{RP}$ color-magnitude diagram for the periodic variables with $\rm Prob>0.9$, that have already been discovered (left), and the new discoveries (right). The points are colored by the period.}
    \label{fig:fig6}
\end{figure*}

The Wesenheit $W_{RP}$ vs. $G_{BP}-G_{RP}$ color-magnitude diagram for all the variables with $\rm Prob>0.9$ and the points colored according to the period is shown in Figure \ref{fig:fig6}.  This essentially highlights the large dynamic range in period probed by the ASAS-SN light curves. Owing to the ASAS-SN survey cadence and our long time baseline, we are able to probe both short period variability ($P<0.1$ d) and long period variability ($P>1000$ d). The ASAS-SN survey continues to monitor the sky in the g-band, which lends itself well to the analysis of long term trends and unusual variability. As a testament to this, \citet{2019ATel12836....1J} noted a sudden dimming episode (flux reduction of ${\sim}70\%$ in the g-band) in an APASS source (ASASSN-V J213939.3$-$702817.4) that was non-variable for ${\sim}1800$ d. This source was classified as a constant source in this work.

\begin{figure*}
	\includegraphics[width=\textwidth]{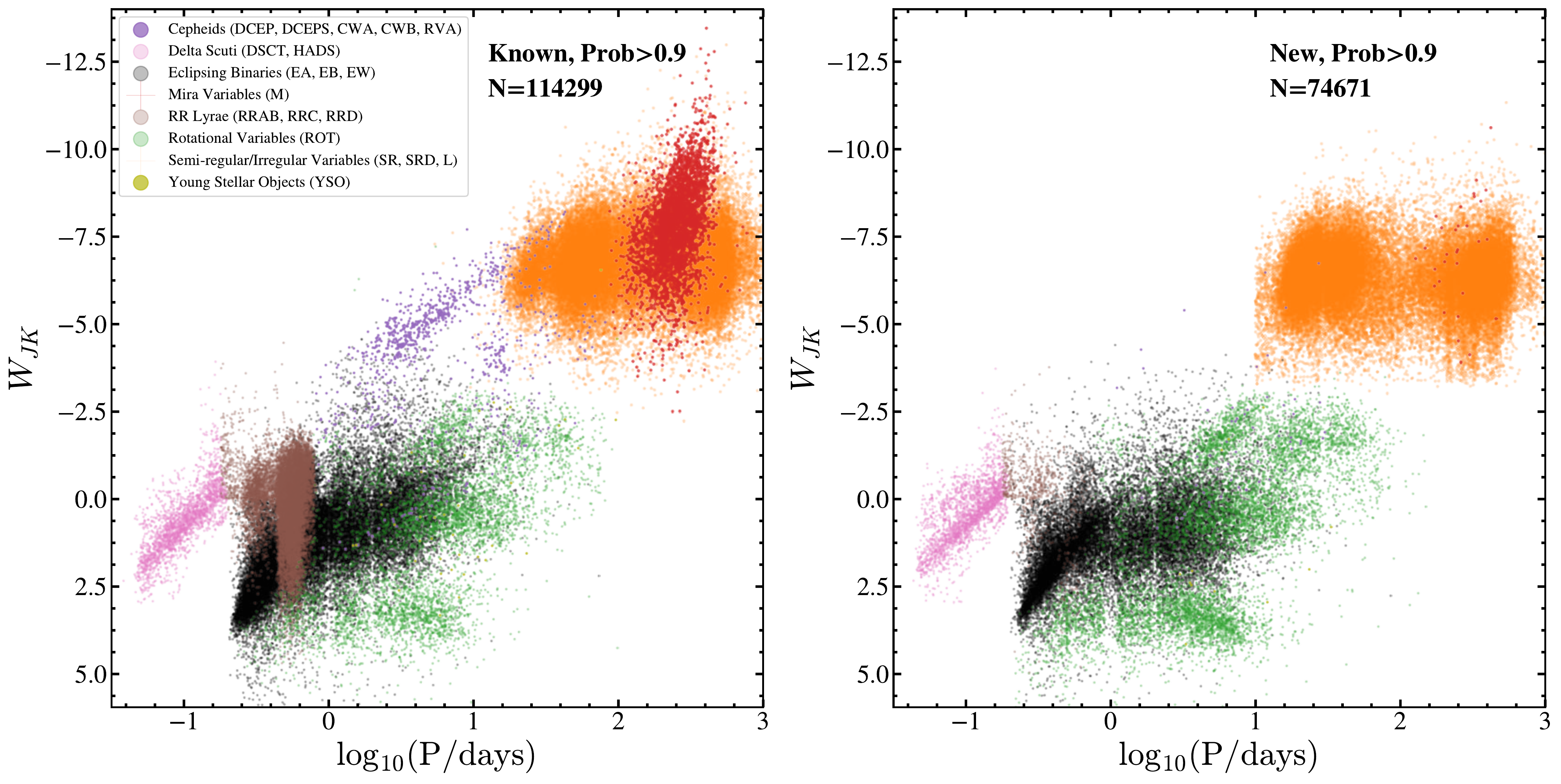}
    \caption{The Wesenheit $W_{JK}$ PLR diagram for the periodic variables with $\rm Prob>0.9$, that have already been discovered (left), and the new discoveries (right).}
    \label{fig:fig7}
\end{figure*}

The combined Wesenheit $W_{JK}$ period-luminosity relationship (PLR) diagram for the periodic variables with $\rm Prob>0.9$ is shown in Figure \ref{fig:fig7}. The PLR sequences for the Cepheids are well defined \citep{2005AcA....55..331S}. Sharp PLR sequences can also be seen for Delta Scuti variables and contact binaries. The Mira variables also form a distinct PLR sequence beyond $P>100$ d. The slight deficits of variables at the aliases of a sidereal day (e.g., $P\approx1$ d, $P\approx2$ d, $P\approx30$ d, etc.) are due to the quality checks implemented in $\S3.5$.

\begin{figure*}
	\includegraphics[width=\textwidth]{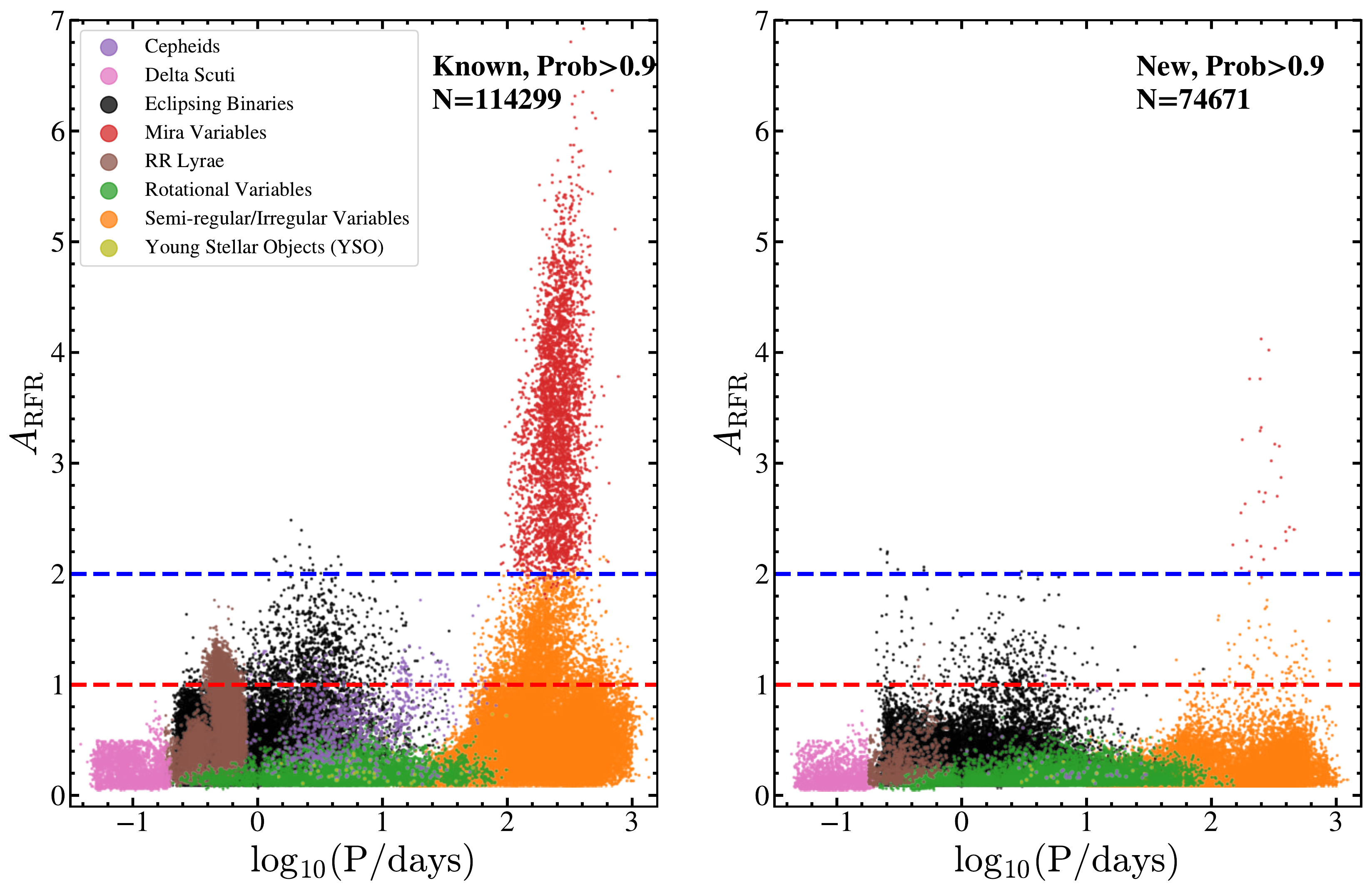}
    \caption{Period-amplitude plot for the for the periodic variables with $\rm Prob>0.9$, that have already been discovered (left), and the new discoveries (right). Reference amplitudes of 1 and 2 mag are shown in red and blue respectively.  }
    \label{fig:fig8}
\end{figure*}

The period-amplitude plot for the periodic variables with $\rm Prob>0.9$ is shown in Figure \ref{fig:fig8}. The high prior completeness of the Mira, RR Lyrae and Cepheid variables is evident. We do not discover many of these variables in this work. The large majority (${\sim}98.7\%$) of the new discoveries are of different variable types with smaller variability amplitudes and/or weak periodicity.

\begin{figure*}
	\includegraphics[width=\textwidth]{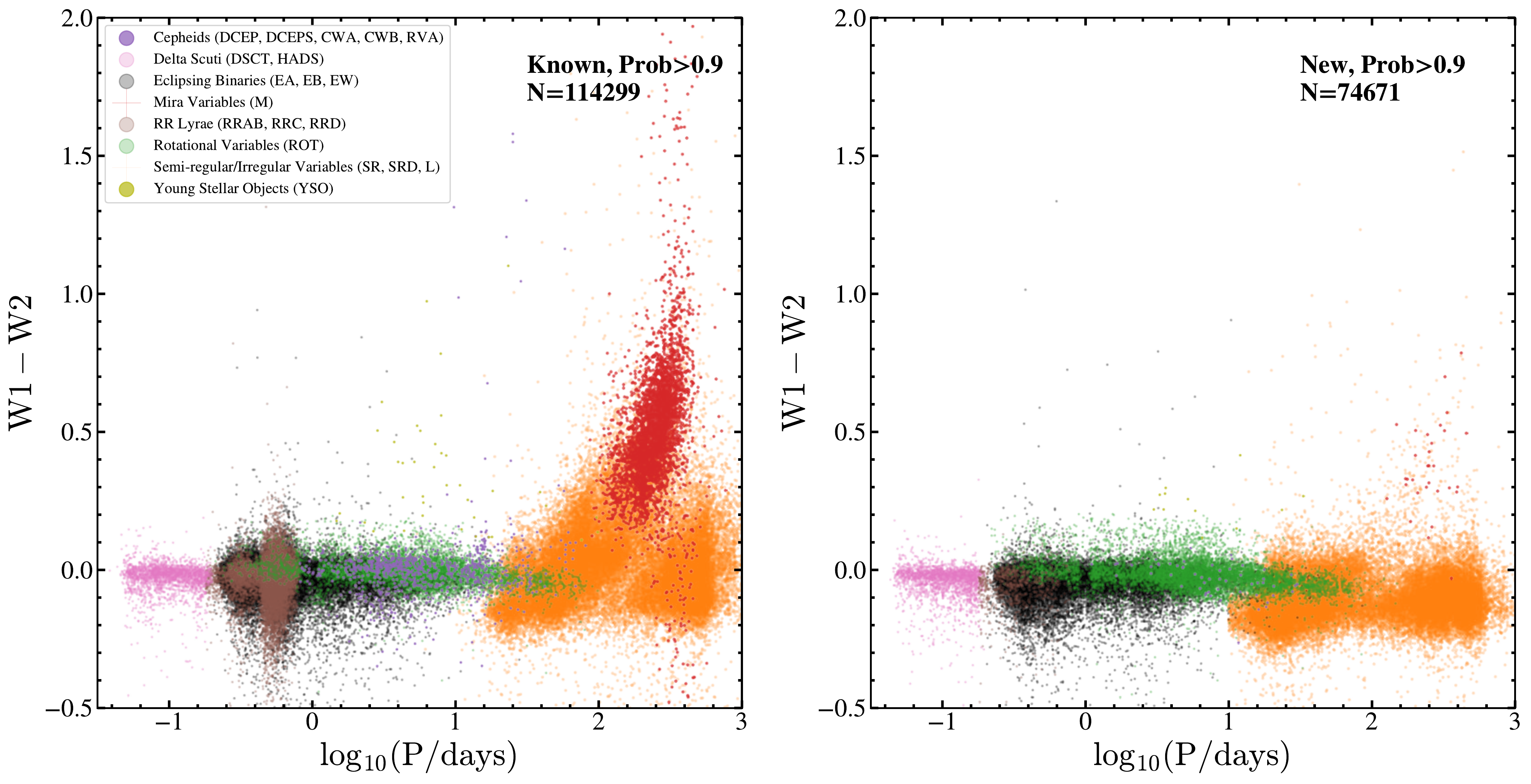}
    \caption{The period vs. $\rm W_1-\rm W_2$ color diagram for the variables with $\rm Prob>0.9$, that have already been discovered (left), and the new discoveries (right). }
    \label{fig:fig9}
\end{figure*}

We also examine the period-color relationship of the variables in the AllWISE \citep{2013yCat.2328....0C,2010AJ....140.1868W} $\rm W_1-\rm W_2$ color space in Figure \ref{fig:fig9}. Most variables have $\rm W_1-\rm W_2$ $\sim0$, but the NIR infrared-excess increases with increasing period for the long period variables. This is even more dramatic for the Mira variables. Dust formation is commonly traced through infrared excesses. Our findings agree with \citet{2018MNRAS.481.4984M}, for example, that strong mass loss and increased dust formation first occurs for pulsation periods of $P{\gtrsim}60$ d for Galactic stars. 

\begin{figure*}
	\includegraphics[width=\textwidth]{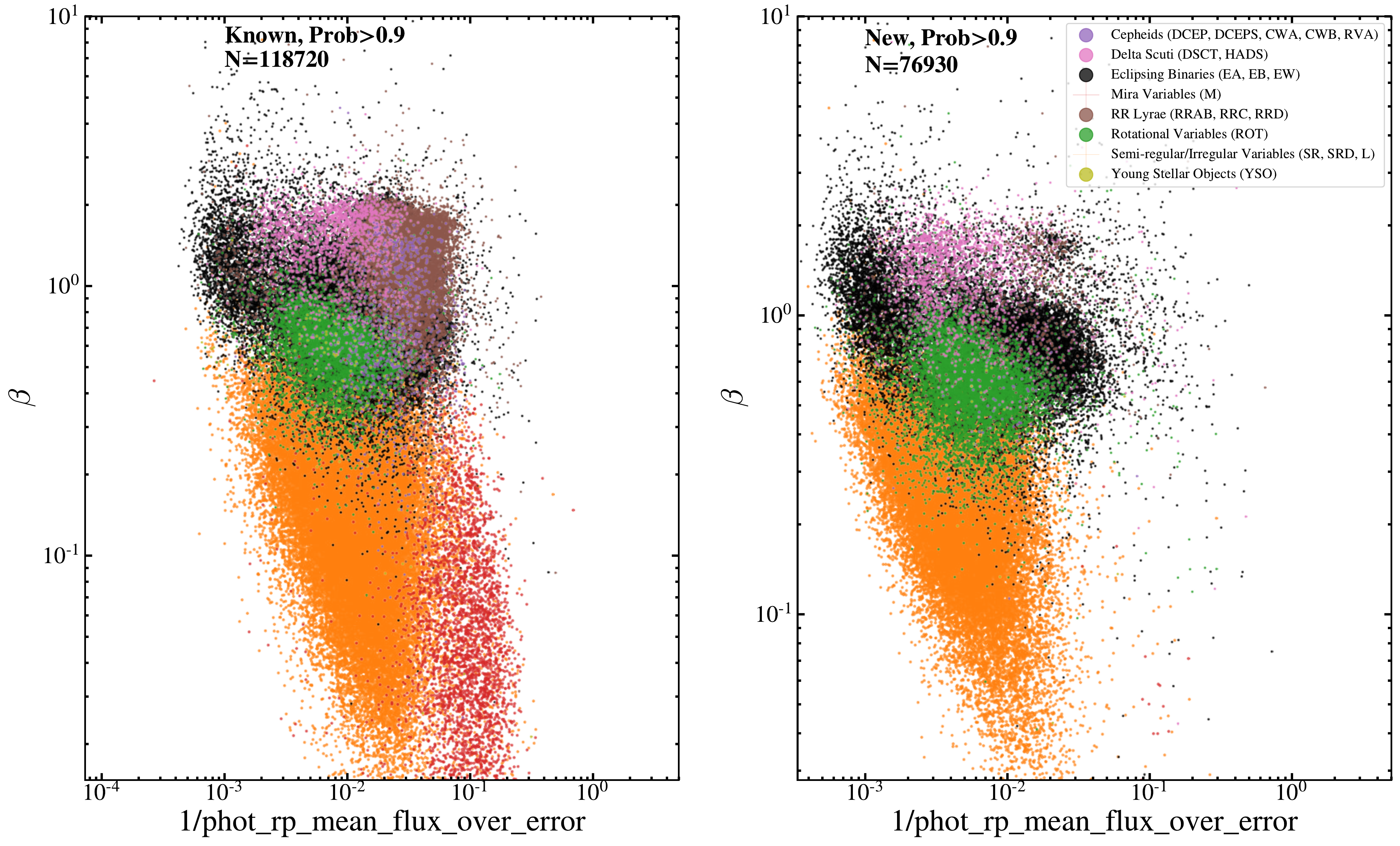}
    \caption{The Gaia DR2 BP/RP variability ratio $\beta$ against 1/$\rm phot\_rp\_mean\_flux\_over\_error$}
    \label{fig:fig10}
\end{figure*}

As an external check of our classifications, we used data from our cross-match to Gaia DR2 \citep{2018arXiv180409365G} to produce Figure \ref{fig:fig10}. We define a ``variability'' color $\beta$,\begin{equation}
    \beta=\rm phot\_bp\_mean\_flux\_error/phot\_rp\_mean\_flux\_error \,, 
	\label{eq:alp}
\end{equation} which is a measure of the difference in variability between the bluer and redder Gaia bands and compare it to the inverse of the quantity \verb "phot_rp_mean_flux_over_error"     which is a measure of the mean signal to noise ratio. The different groups of variables fall in distinct regions, with red pulsating variables having smaller values of $\beta$ compared to bluer variables. Comparing the known variables and the new discoveries, we find that the new discoveries mostly fall in the same regions as the known variables. This provides an independent confirmation of the purity of the newly discovered variables and validates our quality assurance methodology in $\S 3.5$.

\begin{figure*}
	\includegraphics[width=\textwidth]{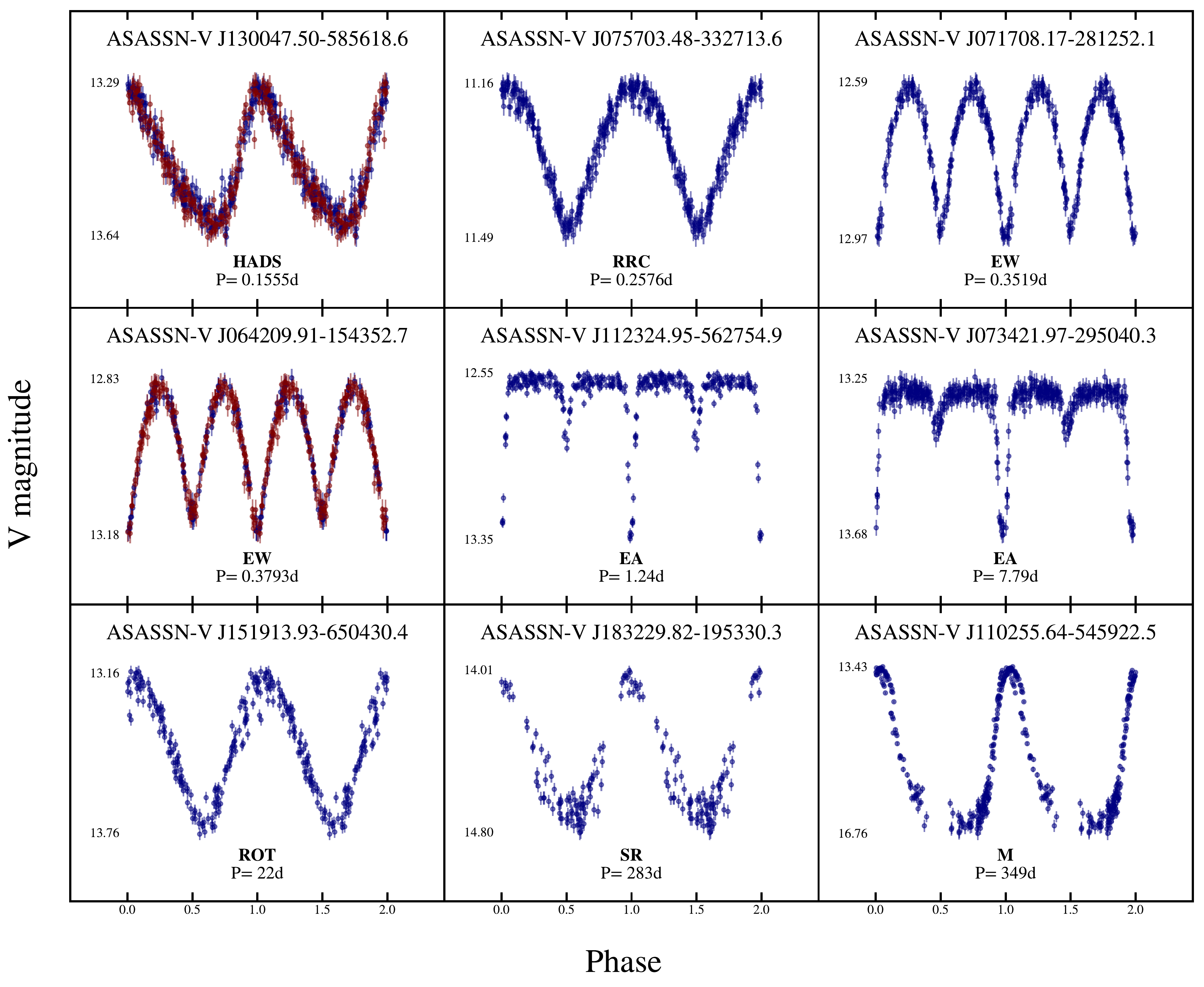}
    \caption{Phased light curves for examples of the newly discovered periodic variables. The light curves are scaled by their minimum and maximum V-band magnitudes. Different colored points correspond to data from the different ASAS-SN cameras. The different variability types are defined in Table \ref{tab:var}.}
    \label{fig:fig11}
\end{figure*}

\begin{figure*}
	\includegraphics[width=\textwidth]{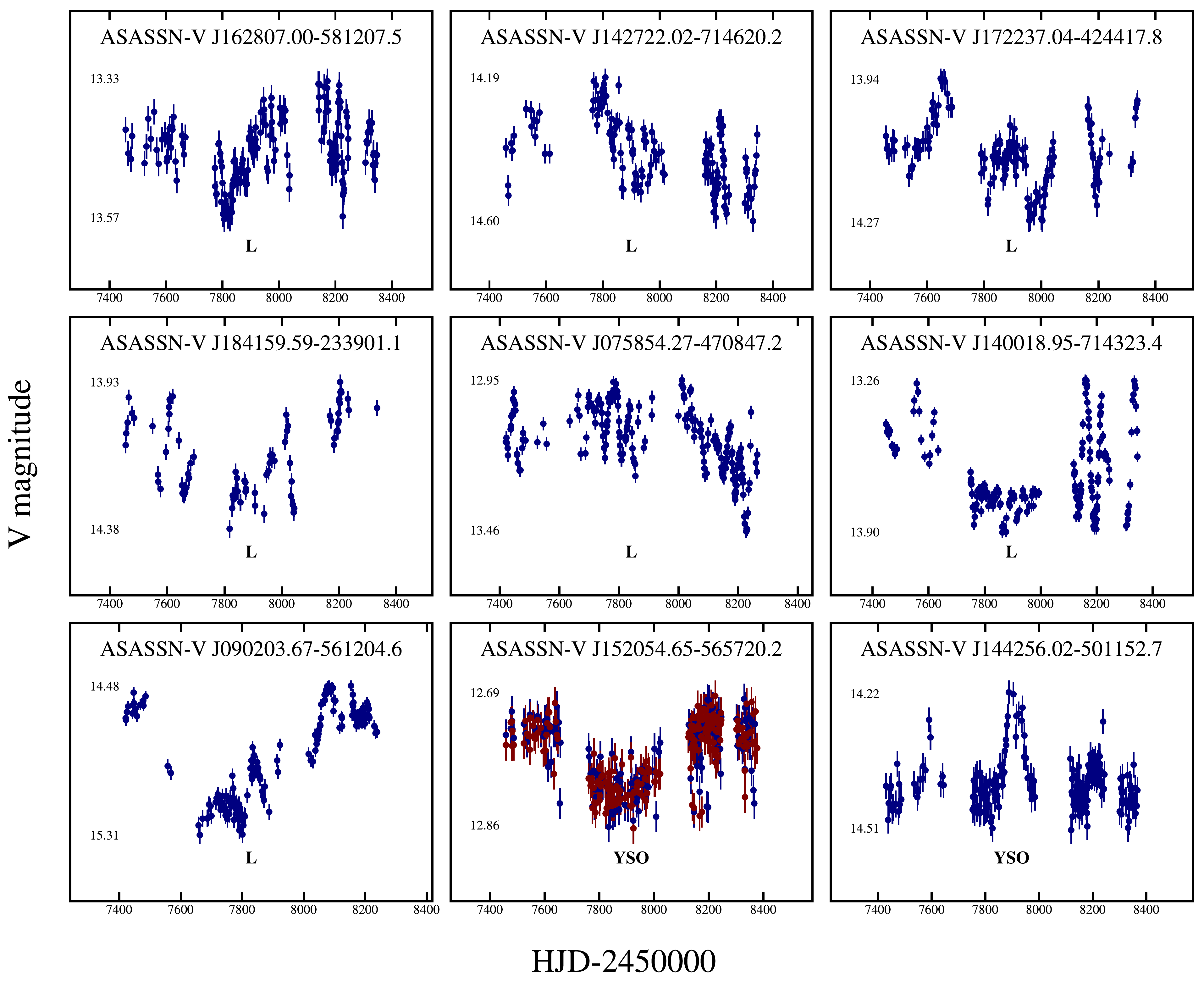}
    \caption{Light curves for examples of the newly discovered irregular variables. The format is the same as for Fig. \ref{fig:fig11}}
    \label{fig:fig12}
\end{figure*}

Examples of the newly identified periodic variables are shown in Figure \ref{fig:fig11} and examples of the newly discovered irregular variables are shown in Figure \ref{fig:fig12}. The light curves for the red giant pulsators, including the irregular variables, are complex, and often multi-periodic, which requires further Fourier analysis. In order to better understand these pulsating red giants, \citet{2019arXiv190503279P} recommended a more detailed analysis, combining visual inspection of the light curves and a more advanced period analysis, in lieu of the automated classification used by ASAS-SN.

\begin{figure*}
	\includegraphics[width=\textwidth]{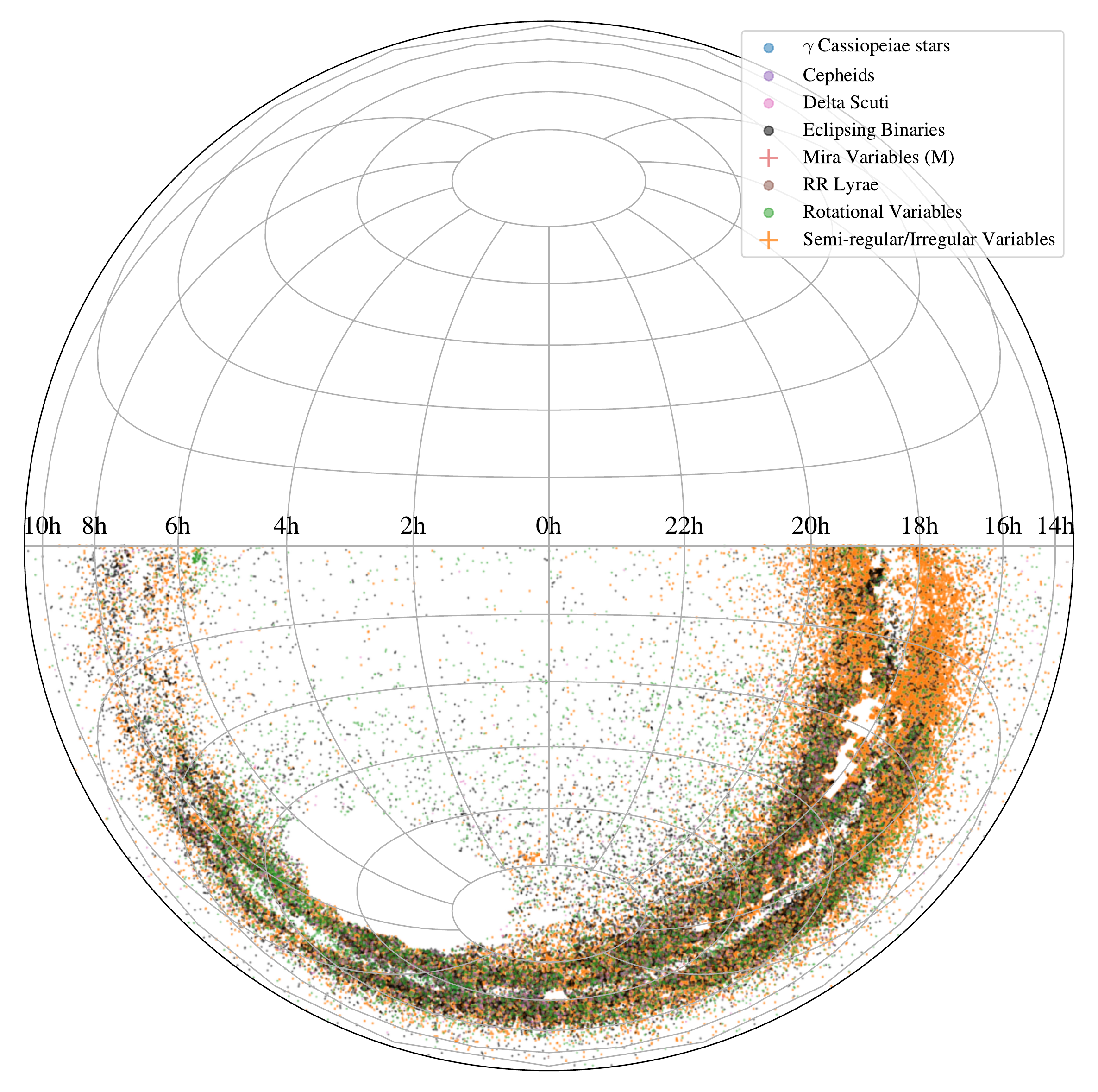}
    \caption{Spatial distribution of the $\sim 88,300$ newly discovered variables in Equatorial coordinates. Sources in the gap centered at the Southern Ecliptic Pole ($\alpha=6$ h, $\delta=-66.55$ deg) were analyzed in \citet{2019MNRAS.485..961J}. The other gaps are in the APASS catalog.}
    \label{fig:fig13}
\end{figure*}

\begin{figure*}
	\includegraphics[width=\textwidth]{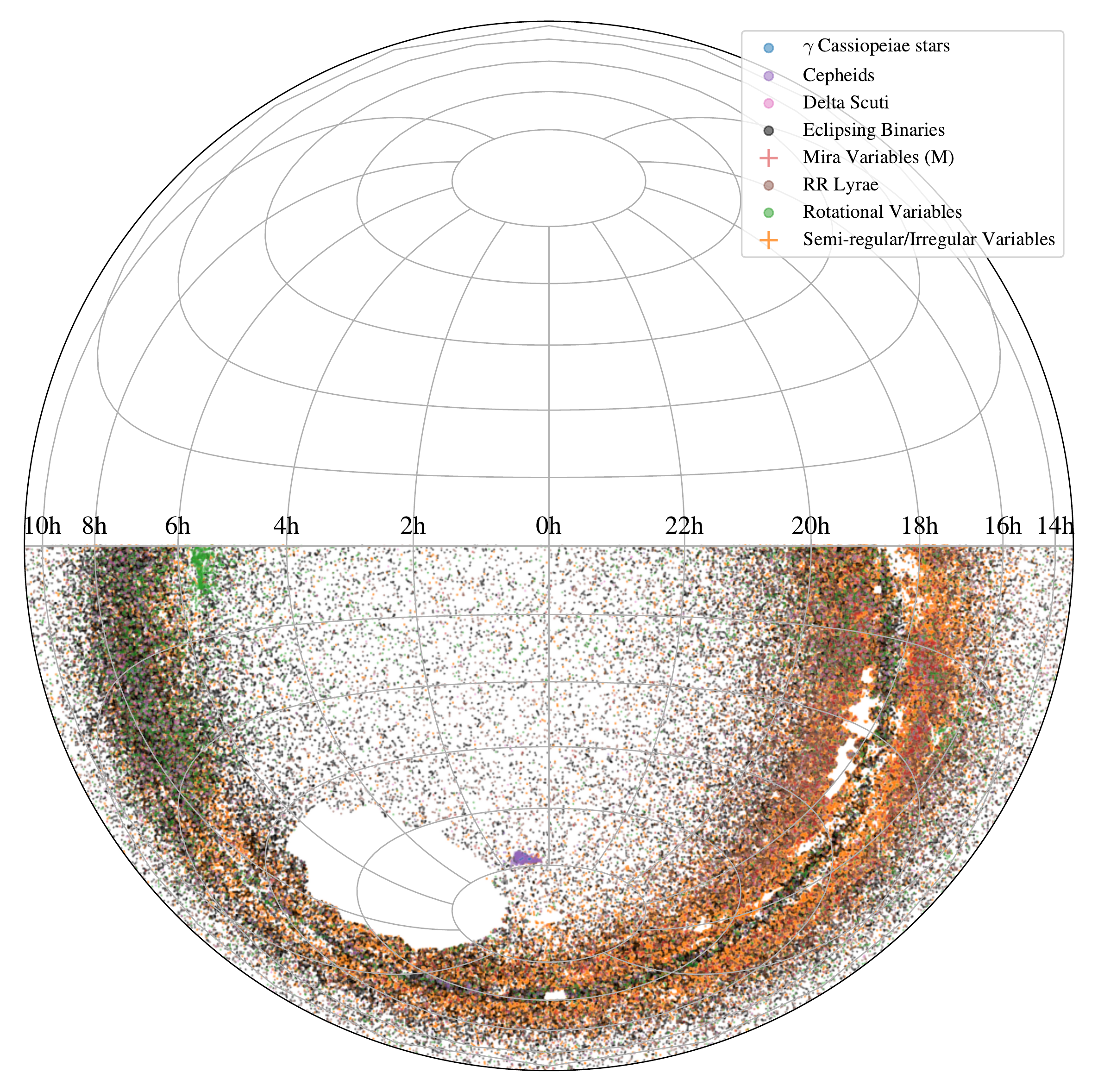}
    \caption{Spatial distribution of the $\sim 131,900$ known variables in Equatorial coordinates. Sources in the gap centered at the Southern Ecliptic Pole ($\alpha=6$ h, $\delta=-66.55$ deg) were analyzed in \citet{2019MNRAS.485..961J}. The other gaps are in the APASS catalog.}
    \label{fig:fig14}
\end{figure*}

We illustrate the sky distribution of the newly discovered variables in Figure \ref{fig:fig13}. Most of these discoveries are clustered towards the Galactic disk, as is expected. We note the scarcity of high amplitude Mira variables, RR Lyrae and Cepheid variables and the abundance of lower amplitude semi-regular/irregular variables among the newly discovered variables. Variables with large amplitudes and strong periodicity are relatively easily discovered and characterized by wide field photometric surveys, so the existing completeness of these variable types is very high. The gaps in coverage will be rectified in the next paper in this series. We also show the sky distribution of the known variables identified in this work in Figure \ref{fig:fig14}. Here, we note the abundance of Mira variables, eclipsing binaries and Cepheid variables that have been discovered by previous surveys.

\section{Conclusions}

We systematically searched for variable sources with $V<17$ mag in the southern hemisphere ($\delta<0$ deg), excluding the ${\sim}1.3$M sources near the Southern Ecliptic Pole which were analyzed in Paper III. Through our search, we identified ${\sim}220,000$ variable sources, of which ${\sim}88,300$ are new discoveries. The sample of new variables includes ${\sim}48,000$ red pulsating variables, ${\sim}23,000$ eclipsing binaries, ${\sim}2,200$ $\delta$-Scuti variables and ${\sim}10,200$ rotational variables.

The V-band light curves of all the ${\sim}30.1$M sources studied in this work are available online at the ASAS-SN Photometry Database (\url{https://asas-sn.osu.edu/photometry}). We have also updated the ASAS-SN variable stars database (\url{https://asas-sn.osu.edu/variables}) with the light curves of these new variables. Most of these sources will also fall into the TESS footprint, thus short baseline TESS light curves that possess better photometric precision can also be obtained to complement the long baseline ASAS-SN light curves. 

This work greatly improves the completeness of bright variables in the Southern hemisphere and provides long baseline V-band light curves. In particular, we have significantly improved the completeness of lower amplitude variables. As part of our ongoing effort to systematically analyze all the ${\sim}50$ million $V<17$ mag APASS sources for variability, we will next update this database with the light curves for the sources across the northern hemisphere and include the light curves for sources missing from the APASS DR9 catalog.

\section*{Acknowledgements}
We thank the anonymous referee for the very useful comments that improved our presentation of this work. We thank Dr. Hans-Walter Rix for his suggestion about using the Gaia DR2 variability information in this analysis. We thank the Las Cumbres Observatory and its staff for its
continuing support of the ASAS-SN project. We also thank the Ohio State University College of Arts and Sciences Technology Services for helping us set up and maintain the ASAS-SN variable stars and photometry databases.

ASAS-SN is supported by the Gordon and Betty Moore
Foundation through grant GBMF5490 to the Ohio State
University and NSF grant AST-1515927. Development of
ASAS-SN has been supported by NSF grant AST-0908816,
the Mt. Cuba Astronomical Foundation, the Center for Cos-
mology and AstroParticle Physics at the Ohio State Univer-
sity, the Chinese Academy of Sciences South America Center
for Astronomy (CAS- SACA), the Villum Foundation, and
George Skestos. 

CSK is supported by NSF grants AST-1515876 , AST-1515927 and AST-1814440. This work is supported in part by Scialog Scholar grant 24216 from the Research Corporation. TAT acknowledges support from a Simons Foundation Fellowship and from an IBM Einstein Fellowship from the Institute for Advanced Study, Princeton. Support for JLP is provided in part by the Ministry of Economy, Development, and Tourism's Millennium Science Initiative through grant IC120009, awarded to The Millennium Institute of Astrophysics, MAS. SD acknowledges Project 11573003 supported by NSFC. Support for MP and OP has been provided by INTER-EXCELLENCE grant LTAUSA18093 from the Czech Ministry of Education, Youth, and Sports. The research of OP has also been supported by Horizon 2020 ERC Starting Grant ``Cat-In-hAT'' (grant agreement \#803158) and PRIMUS/SCI/17 award from Charles University. This work was partly supported by NSFC 11721303.

This work has made use of data from the European Space Agency (ESA)
mission {\it Gaia} (\url{https://www.cosmos.esa.int/gaia}), processed by
the {\it Gaia} Data Processing and Analysis Consortium. This publication makes 
use of data products from the Two Micron All Sky Survey, as well as
data products from the Wide-field Infrared Survey Explorer.
This research was also made possible through the use of the AAVSO Photometric 
All-Sky Survey (APASS), funded by the Robert Martin Ayers Sciences Fund. 

This research has made use of the VizieR catalogue access tool, CDS, Strasbourg, France. 
This research also made use of Astropy, a community-developed core Python package for 
Astronomy (Astropy Collaboration, 2013).








\bsp	
\label{lastpage}
\end{document}